\documentclass[10pt,aps,prd,showpacs,superscriptaddress,twocolumn]{revtex4}
\usepackage{amsfonts}
\usepackage{amssymb}
\usepackage{amsmath}
\usepackage{MnSymbol}
\usepackage{graphicx}

\date{\today}

\begin{document}

\title{Relativistic spring-mass system}

\author{Rodrigo Andrade e Silva}
\email{rasilva@terpmail.umd.edu}
\affiliation{Department of Physics, University of Maryland,
College Park, MD 20742-4111, USA}

\author{Andr\'e G. S. Landulfo}
\email{andre.landulfo@ufabc.edu.br}
\affiliation{Centro de Ci\^encias Naturais e Humanas,
Universidade Federal do ABC, 
Avenida dos Estados, 5001, 09210-580, 
Santo Andr\'e, S\~ao Paulo, Brazil}

\author{George E. A. Matsas}
\email{matsas@ift.unesp.br}
\affiliation{Instituto de F\'\i sica Te\'orica, Universidade Estadual Paulista,
Rua Dr. Bento Teobaldo Ferraz 271, 01140-070, S\~ao Paulo, S\~ao Paulo, Brazil}

\author{Daniel A. T. Vanzella}
\email{vanzella@ifsc.usp.br}
\affiliation{Instituto de F\'\i sica de S\~ao Carlos,
Universidade de S\~ao Paulo, Caixa Postal 369, 13560-970, 
S\~ao Carlos, S\~ao Paulo, Brazil}

\pacs{03.30.+p,03.50.-z}

\begin{abstract}
The harmonic oscillator plays a central role in physics describing the dynamics of a wide 
range of systems close to stable equilibrium points. The nonrelativistic one-dimensional 
spring-mass system is considered a prototype representative of it. It is usually assumed 
and galvanized in textbooks that the equation of motion of a relativistic harmonic oscillator 
is given by the same equation as the nonrelativistic one with the mass $M$, at the tip, 
multiplied by the relativistic factor $1/(1 - v^2/c^2)^{1/2}$. Although the solution of such 
an equation may depict some physical systems, it does not describe, in general, 
one-dimensional {\em relativistic} spring-mass oscillators under the influence of elastic 
forces. In recognition to the importance of such a system to physics, we fill a gap in the 
literature and offer a full relativistic treatment for a system composed of a spring attached 
to an inertial wall, holding a mass  $M$ at the end.
\end{abstract}

\maketitle

\section{Introduction}
\label{sec:introduction}

The harmonic oscillator has universal importance to physics, 
once it describes small oscillations close to stable equilibrium 
points of general systems. A prototype harmonic oscillator is the 
nonrelativistic one-dimensional spring-mass system, since the 
parabolic potential is approximately valid (for a variety 
of materials) as far as the (small-mass) spring does not suffer 
plastic deformations. In addition, it has been assumed for long and 
galvanized in textbooks (see, e.g., Refs.~\cite{pz,h,klml,gps} and references 
therein) that the motion $x=x(t)$ of a {\em relativistic} 
``harmonic'' oscillator would satisfy
\begin{equation}
\frac{d}{dx^0} \frac{M\dot x}{(1-\dot x^2/c^2)^{1/2}} 
= - \frac{\partial \Phi}{\partial x}, 
\label{wronghooke1}
\end{equation}    
where  $M$ is the particle (rest) mass, $c$ is the speed of light,
$'\;\dot{} \; ' \equiv d/dx^0$, $x=x^1$, and $x^\mu$ 
are usual inertial Cartesian coordinates covering the Minkowski spacetime 
$(\mathbb{R}^4,\eta_{\mu \nu})$ with metric $\eta_{\mu \nu}$ chosen here
to have signature $(+,-,-,-)$. The potential corresponding to Hooke's 
law would be $\Phi = kx^2/2$, where $k$ is the usual elastic 
constant. Correspondingly, the energy conservation equation would 
be given by
\begin{equation}
\frac{M c^2}{(1-\dot x^2/c^2)^{1/2}} + \Phi \equiv {\cal E} = {\rm const}.
\label{wronghooke2}
\end{equation}
It should be stressed, however, that although Eqs.~(\ref{wronghooke1}) 
and~(\ref{wronghooke2}) may describe some physical systems 
(see Appendix~\ref{App:Appendix}), {\em they do not characterize
one-dimensional relativistic spring-mass oscillators under the influence of 
elastic forces}. 

Although some attention has been devoted to the issue
of elasticity in the relativistic context~\cite{bs,bp}, it 
is somewhat surprising that the relativistic spring-mass system 
has not received a comprehensive relativistic treatment yet. 
This may be because the speed of sound $v_s$ is typically 
small under usual conditions, e.g., approximately 13~km/s for 
beryllium and silicon carbide, 12~km/s for diamond and 10~km/s 
for aramid (a class of synthetic fiber), while maximum local velocities 
$v_m$ should be even smaller. (As it will be seen further, $v_m$ 
may be estimated by multiplying $v_s$ by the maximum strain, i.e., the 
maximum relative spring deformation.)   
In spite of the small local velocity rates achieved in most 
materials under normal conditions, we must recall that this may be 
radically altered under extreme regimes. In neutron
stars, e.g., the sound and light velocities can become comparable.  

In view of the importance of harmonic oscillators to 
physics, we offer here a comprehensive treatment of the 
relativistic one-dimensional Hookean spring-mass system in Minkowski 
spacetime. The paper is organized as follows. In 
Sec.~\ref{sec:Tmunu}, we define the spring stress-energy-momentum 
tensor $T_{ab}$. In Sec.~\ref{sec:spring dynamics}, we 
use $T_{ab}$ derived earlier to describe the spring dynamics.  
In Sec.~\ref{sec:causality}, we impose conditions on the spring 
to respect causality and show that the equations of motion are 
hyperbolic if and only if the weak energy condition is satisfied. 
In Sec.~\ref{sec:boundary conditions}, we define the boundary conditions
and discuss energy conservation. In Secs.~\ref{sec:nonrelativistic} 
and~\ref{sec:relativistic}, we apply our results to discuss the 
nonrelativistic and relativistic spring-mass systems, respectively. 
Conclusions are presented in Sec.~\ref{sec:discussions}. We
assume units where $c=1$, unless stated otherwise, and metric signature
$(+\, -\, -\, -)$.

\begin{figure}
\includegraphics[scale=0.3]{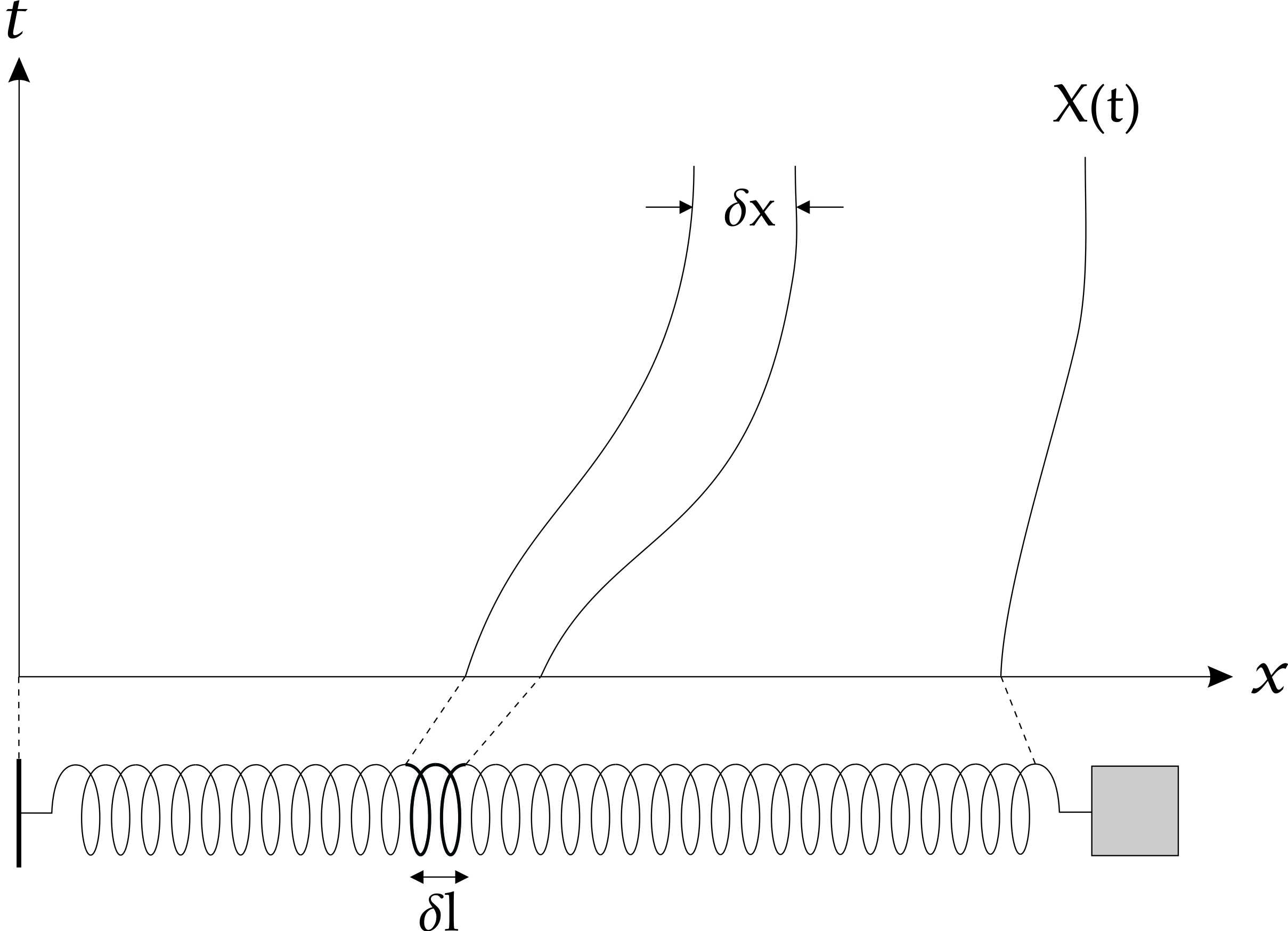}
\caption{The spring points are labeled by their 
(relaxed) proper distance $l$ to the wall chosen 
to lie at rest at the Cartesian coordinate $x^1=0$. 
The Cartesian coordinates of point $l$ at moment $x^0 \equiv t$ 
are $x^1 = x (t,l)$, $x^2=x^3=0$.}
\label{fig1:EC}
\end{figure}
 
\section{The stress-energy tensor}
\label{sec:Tmunu}

Let us consider a one-dimensional spring-mass system in Minkowski 
spacetime set to oscillate at relativistic velocities. 
The spring is assumed to be fixed to an infinitely massive 
inertial wall at one end and to a point mass $M$ at the 
other one. The relaxed spring mass and proper length will be 
denoted by $m$ and $L$, respectively. We might feel 
tempted to neglect the spring mass when $m\ll M$ by imposing $m=0$ 
from the beginning (as it is usual in Newtonian physics). However, 
this will be proved illegal as far as one requests the equation 
of motion to be hyperbolic and the weak energy condition to be 
satisfied. Furthermore, we shall see that contributions due to 
the spring mass are typically larger than the ones due to 
relativistic corrections. Thus, we would rather make no assumptions 
on $m/M$ (except in Sec.~\ref{sec:nonrelativistic}, 
where we expand the equation of motion in terms of $m/M \ll 1$ 
in order to expose more clearly the physical content of our results). 

Our spring will obey Hooke's law in the sense that the proper force 
at the tip of the {\em static} spring will be $-k (X-L)$, where $k$ is the 
spring constant and $X$ is the compressed/stretched spring proper length.
It is also assumed to be homogeneous: 
not only it will have constant linear density $\lambda = m/L$ along 
its length, but if we cut out some portion of it with relaxed 
proper length $L'$, such a portion  will behave as a spring 
with $k' = k (L/L')$ itself. For the sake of simplicity, 
our spring will be modeled as a long cylinder with cross 
sectional area $a$ and vanishing Poisson's ratio (i.e., $a={\rm const}$ 
during the motion) in which case the Young's modulus is 
$Y \equiv k L/a = {\rm const}$. 

We cover the Minkowski spacetime with inertial Cartesian coordinates $x^\mu$ 
and set the wall to lie at the origin. We also let the $x^1$ axis to be aligned 
with the spring motion. Each spring point will be labeled by a parameter, 
$l \in \mathbb{R}$, corresponding to the proper distance between the 
point and the wall when the spring is relaxed. We denote by $x^1 = x(t,l)$ 
the $x^1$ coordinate of the point $l$ at time $t$. The coordinates $(t,l)$ 
are related to the inertial Cartesian coordinates $(x^0,x^1)$ by the transformation
\begin{eqnarray}\label{CT}
x^0=t, \;\;\; \;\;\;  x^1=x(t,l).
\end{eqnarray}
The position of the mass $M$ attached to the end of the spring 
(see Fig.~\ref{fig1:EC}) is given by $X(t)=x(t,L)$.  

In order to derive the equation of motion, we should 
first determine the spring's stress-energy tensor 
$T^{ab}$. Inertial observers instantaneously at rest 
with respect to some arbitrary spring point $l$ may adapt
a local inertial Cartesian coordinate system $ {x'}^\mu= (t',x',y',z')$
and define 
$
T \equiv T^{\mu \nu}_0 \partial/ \partial {x'}^\mu \otimes \partial/ \partial {x'}^\nu,
$  
with    
\begin{equation}\label{T0}
T^{\mu \nu}_0 = \left(
\begin{array}{cccc}
\rho_0 & 0 & 0 & 0 \\
0 & \tau_0 & 0 & 0 \\
0 & 0 & 0 & 0 \\
0 & 0 & 0 & 0
\end{array} \right),
\end{equation}
where $\rho_0$ and $\tau_0$ are the proper mass density and 
pressure along the $x$-direction, respectively. We note that the 
one-dimensional character of the problem, together with the 
negligible Poisson ratio of the spring should prevent any other tension 
components to appear in Eq.~(\ref{T0})~\cite{note1}. 
We will assume that according to such instantaneously comoving observer, 
close enough points at $l + \delta l$ move sufficiently slowly to be 
under the influence of Hooke's force $F$. (Up to first order in 
$\partial (\partial x/ \partial l)/\partial t $, this assumption  is 
not any stronger than the one required for Newtonian springs.) 
As a result, we can write 
$\tau_0 \equiv F/a= - Y (\gamma \delta x - \delta l)/\delta l$ as
\begin{equation}\label{HL}
\tau_0 =  - Y \left( \gamma \frac{\partial x}{\partial l} - 1 \right),
\end{equation}
 where 
$
\delta x \equiv x(t,l+\delta l) - x(t,l)
$
and
$\gamma \equiv (1-u^2)^{-1/2}$,  with 
$u\equiv {\partial x(t,l)}/{\partial t}.$ 
The factor $\gamma$ was introduced to convert the coordinate
distance $\delta x$ into proper distance. We recall that because 
of the spring homogeneity, $Y=kL/a=k' \delta l/a $, with $k'$ 
being the spring constant of a relaxed segment with
proper size  $\delta l$.  

\begin{figure}[h]
\includegraphics[scale = 0.35]{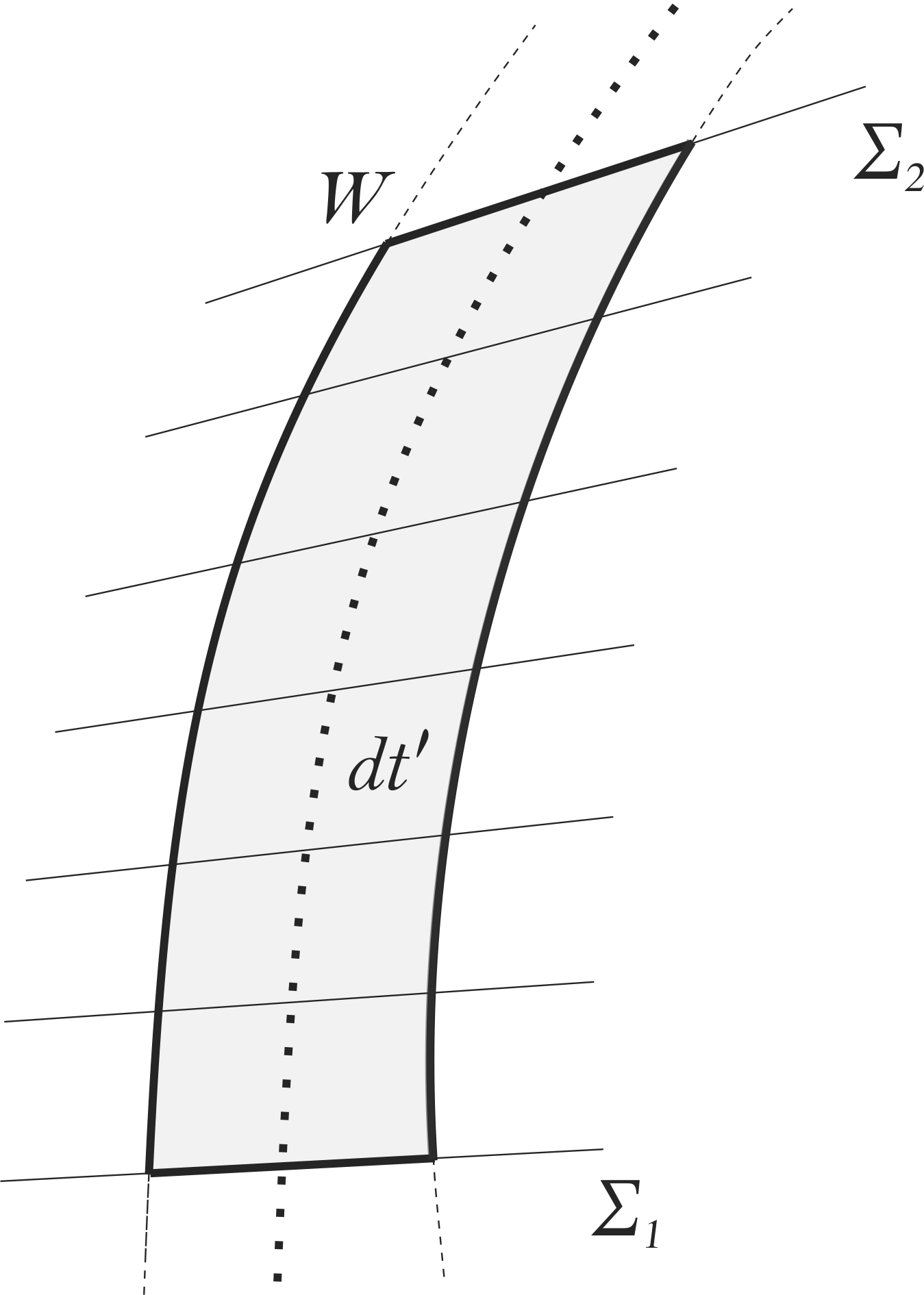}
\caption{The region $W,$ outlined with bold lines, is the portion 
of the worldtube of the spring containing the points in the interval  
$\mathcal{I}=(l, l+\delta l)$. This region is sliced by a family of 
spacelike hypersurfaces $\Sigma_{t'}$ orthogonal to the worldline 
of some fiducial spring point in the interval and labeled by its proper
 time $t'$. }
\label{fig2:W}
\end{figure}

Since there are no external forces acting on the interior points of the spring, the stress energy
tensor satisfies 
\begin{equation}\label{divT}
\nabla_a T^{ab}=0
\end{equation}
for $0<l<L$.
This yields two independent equations, which should be solved for $x(t,l)$ and 
$\rho_0(t,l)$. In order to obtain $\rho_0$, let us use Stokes' theorem
and Eq.~(\ref{divT}) to write
\begin{equation}
\int_W T^{ab} \nabla_a U_b \, d\Xi = \int_{\partial W} T^{ab}U_b \, d \Sigma_a.  
\label{intT0}
\end{equation}
Here, $U^a$ is the 4-velocity vector field of the spring points, $W$ is the spacetime region 
defined by the worldlines in the interval $\mathcal{I}\equiv(l, l+\delta l)$ and delimited 
by the past, $\Sigma_1$, and future, $\Sigma_2$, spacelike hypersurfaces, $\partial W$ 
denotes the boundary of $W$, $d\Xi$ is the spacetime volume element, 
and  $d\Sigma^a$ is the  vector-valued volume element induced
on $\partial W$ (see Fig.~\ref{fig2:W}). Since there is no flux of momentum 
seen by an observer accompanying the matter flow, we can cast Eq.~(\ref{intT0}) as 
\begin{equation}
\delta m_2 - \delta m_1 =  \int_W T^{ab} \nabla_a U_b \, d\Xi,
\label{delta_m}
\end{equation}
where 
\begin{equation}
\delta m_j \equiv  (-1)^{j} \int_{\Sigma_j} T^{ab} U_b \, d\Sigma_ a, \;\; j=1,2,
\end{equation}
is the proper mass of the piece $\mathcal{I}$ of the spring at $\Sigma_j$. (The factor
$(-1)^{j}$ appears only to fix the orientation of $d\Sigma^a$ introduced in
Eq.~(\ref{intT0}), which is
past directed on $\Sigma_1$.)

In order to evaluate the right-hand side of Eq.~(\ref{delta_m}), we will first 
choose some fiducial worldline associated with a point within the  the portion 
$\mathcal{I}$ of the spring. Then,  we will partition the region $W$ into
infinitesimal cells by slicing it with spacelike hypersurfaces $\Sigma_{t'}$ 
(with $\Sigma_{t'_1}\equiv\Sigma_1$ and $\Sigma_{t'_2}\equiv \Sigma_2$) 
differing by an infinitesimal proper time $dt'$ and which are orthogonal to the 
fiducial worldline, as it is shown in Fig.~\ref{fig2:W}. Hence, we can write 
\begin{equation}
\int_W T^{ab} \nabla_a U_b \, d\Xi = 
\int_{t'_1}^{t'_2} dt' \int_{\Sigma_{t'}} dx'dy'dz'  T^{ab} \nabla_a U_b.
\label{intT0b}
\end{equation}
Here, $(t',x',y',z')$  are local inertial Cartesian coordinates associated, within each cell, 
with the fiducial worldline. In these coordinates, the components of $T^{ab}$ are 
given by $T^{\mu \nu}_0$ in Eq.~(\ref{T0}) and  $U^a$ decomposes as 
$U^\mu =  \gamma' (1, \vec{u}\, ')$, the prime indicating velocities  with 
respect to the local inertial frame being used. As a consequence, 
$u' \ll 1$, $\gamma' \approx 1$, and 
\begin{equation}
T^{ab} \nabla_a U_b 
= T^{00}_0 \partial U_0/\partial t' + T^{11}_0 \partial U_1/\partial x' 
\approx - \tau_0 \frac{\partial {u'}^1}{\partial x'},
\label{T0nablaU}
\end{equation}
with the approximation becoming exact as $\delta l\rightarrow 0$. 
Now, if we use Eq.~(\ref{T0nablaU}) together with  the identity 
\begin{equation}
\frac{d}{dt'}\delta V' = (\nabla \cdot \vec u' ) \delta V',
\end{equation}
where $\delta V'\equiv a \delta x' $ and  $\delta x'$ is the length of the 
piece of the spring as measured in the local inertial frame, we can cast 
Eq.~(\ref{intT0b}) as 
\begin{equation}
\int_W T^{ab} \nabla_a U_b \, d\Xi = -\int_{t'_1}^{t'_2} dt' a \tau_0 \frac{d (\delta x')}{dt'},
\end{equation}
where $\int_{\Sigma_{t'}}$ was suppressed once we are looking at a spring element.
By making use of the above result in Eq.~(\ref{delta_m}), we find that the change in the 
proper mass of the piece of the spring as it evolves from $\Sigma_1$ to $\Sigma_2$  is    
\begin{equation}
\delta m_2 - \delta m_1
= - \int_{t'_1}^{t'_2} dt' a \tau_0 \frac{d (\delta x')}{dt'} 
= - \int_{\delta x'_1}^{\delta x'_2} a \tau_0 d(\delta x').
\end{equation}
Consequently, we conclude that any change in the  mass of the portion~$\mathcal{I}$ 
of the spring is due to the total work done on it by the neighboring matter. Using 
Hooke's law~(\ref{HL}) and integrating from the moment the piece is relaxed,
 where $\delta x'_1 = \delta l$ and $\delta m_1 = \lambda \delta l$, to some 
 arbitrarily time in the wall's frame, where $\delta x'_2= \gamma \delta x$, we find that
\begin{eqnarray}
\delta m_2 - \lambda \delta l 
&=& \frac{aY}{\delta l} \int_{\delta l}^{\gamma \delta x}  (\delta x' - \delta l)d(\delta x') 
\nonumber \\
 &=& \frac{aY}{2\delta l} (\gamma \delta x - \delta l)^2
 \end{eqnarray}
 and hence, 
 \begin{equation}\label{m2}
 \delta m_2 - \lambda \delta l  
 = \frac{aY \delta l}{2} {\left( \gamma \frac{\partial x}{\partial l} - 1 \right)}^2.
\end{equation}
By making use of Eq.~(\ref{m2}), it is possible to write the proper mass density 
$\rho_0 \equiv \delta m_2 / (a \gamma \delta x)$ in terms of $x(t, l)$ as 
\begin{equation}
\rho_0 
= \frac{\lambda}{a} {\left( \gamma \frac{\partial x}{\partial l} \right) \!}^{-1} 
 + \frac{Y}{2} {\left( \gamma \frac{\partial x}{\partial l} \right) \!}^{-1} 
 {\left( \gamma \frac{\partial x}{\partial l} - 1 \right)}^2.
\label{rho0}
\end{equation}

To find the components, $T^{\mu \nu}$, of the stress-energy tensor in the 
frame of the wall, we can simply apply a boost with velocity $(-u,0,0)$ on 
$T^{\mu \nu}_0$ to find that
\begin{equation}\label{T}
T^{\mu \nu} = \left(
\begin{array}{cccc}
\rho & g & 0 & 0 \\
g & p & 0 & 0 \\
0 & 0 & 0 & 0 \\
0 & 0 & 0 & 0
\end{array} \right),
\end{equation}
where
\begin{equation}\label{rho}
\rho = \gamma^2 ( \rho_0 + u^2 \tau_0),
\end{equation}
\begin{equation}\label{g}
g = u \gamma^2 ( \rho_0 + \tau_0),
\end{equation}
and
\begin{equation}\label{p}
p = \gamma^2 (u^2 \rho_0 + \tau_0),
\end{equation}
with $\tau_0$ and $\rho_0$ being given by Eqs.~(\ref{HL}) and~(\ref{rho0}), 
respectively. Hence, we have written the spring's stress-energy tensor entirely in 
terms of the function $x(t,l)$ (and its derivatives), which describes the configuration 
of the spring's points at each time.

\section{Spring dynamics}
\label{sec:spring dynamics}

We want to find now the equation describing the dynamics of $x(t, l)$. 
For this purpose, we will first apply  Eq.~(\ref{T}) into Eq.~(\ref{divT})  in order to obtain 
\begin{equation}\label{divT1}
\frac{\partial \rho}{\partial x^0}+ \frac{\partial g}{\partial x^1} = 0
\end{equation}
and
\begin{equation}\label{divT22}
\frac{\partial g}{\partial x^0}+ \frac{\partial p}{\partial x^1} = 0.
\end{equation}
As we have seen in Sec~\ref{sec:Tmunu}, the quantities $\rho,$ $g$, and
 $p$ can be written in terms of $x(t,l)$ and, therefore, it will be more 
 convenient to rewrite Eqs.~(\ref{divT1}) and~(\ref{divT22}) using the 
 derivatives $\partial/\partial t$ and $\partial/\partial l$. This can be 
 done by making use of the coordinate transformation~(\ref{CT}) to write
\begin{eqnarray}
\frac{\partial}{\partial  x^0} 
&=& - \frac{\partial x}{\partial t} {\left( \frac{\partial x}{\partial l} \right) \!}^{-1}
\frac{\partial}{\partial l} + \frac{\partial}{\partial t}, \label{divTleta1} \\
\frac{\partial}{\partial  x}
 &=& {\left( \frac{\partial x}{\partial l} \right) \!}^{-1} \frac{\partial}{\partial l}. 
\label{divTleta2}
\end{eqnarray}
As a result, Eqs.~(\ref{divT1}) and~(\ref{divT22}) transform into
\begin{equation}\label{EnergyCont}
\frac{\partial x}{\partial l} \frac{\partial \rho}{\partial t} 
- u \frac{\partial \rho}{\partial l}+ \frac{\partial g}{\partial l} = 0
\end{equation}
and
\begin{equation}\label{MomentumCont}
\frac{\partial x}{\partial l} \frac{\partial g}{\partial t} 
- u \frac{\partial g}{\partial l}+ \frac{\partial p}{\partial l} = 0,
\end{equation}
respectively. By using the explicit expressions for $\rho$, $g$,
and $p$ in terms of $x(t,l)$, given by Eqs.~(\ref{rho})-(\ref{p}) 
together with Eqs.~(\ref{HL}) and~(\ref{rho0}), we find that 
Eq.~(\ref{EnergyCont}) is simply   Eq.~(\ref{MomentumCont}) 
multiplied by $u = \partial x/\partial t$. Hence, the only relevant 
equation obtained  for $x(t, l)$ is 
\begin{equation}\label{UE}
\zeta[x(t,l)] \frac{\partial^2 x}{\partial t^2} -  
\gamma^{-2}\frac{\partial^2 x}{\partial l^2} - 
2 \frac{\partial x}{\partial t} \frac{\partial x}{\partial l} 
\frac{\partial^2 x}{\partial l \partial t} = 0,
\end{equation}
where
\begin{equation}\label{zeta}
\zeta[x(t, l)] \equiv \frac{\lambda}{kL} + \frac{1}{2} - 
\gamma^2 \left( \frac{1}{2} + {\left( \frac{\partial x}{\partial t} \right) }^2 \right)  
{\left( \frac{\partial x}{\partial l} \right)}^2 
\end{equation}
and we recall that $\gamma =(1 - ( {\partial x}/{\partial t})^2)^{-1/2}.$ 
The spring dynamics is governed by Eq.~(\ref{UE}). In the next section, we shall
analyze when it determines the spring evolution once initial (at $t = 0$) and boundary
(at $l=0$ and $l=L$) conditions are given.

\section{Causality, hyperbolicity, and the weak energy condition}
\label{sec:causality}

The spring equation, Eq.~(\ref{UE}), is a quasi-linear second order partial 
differential equation. As a result, its classification in hyperbolic, parabolic, 
or elliptic depends on the solution: one needs to take a particular solution 
$x(t,l)$, insert it into the coefficients of the second order derivatives
and then classify it as a linear equation. This corresponds to the evaluation 
of the sign of the quantity
\begin{equation}
\Delta \equiv  \gamma^4 u^2  s^2 +  \gamma^2 \zeta[x(t,l)],
\label{Delta}
\end{equation}
where $s\equiv {\partial x}/{\partial l}$.
With the aid of Eq.~(\ref{zeta}), Eq.~(\ref{Delta}) can be cast as
\begin{equation}
\Delta= \gamma^2  \left[ \frac{\lambda}{kL} 
+ \frac{1}{2} \left( 1 - ( \gamma s )^2 \right) \right].
\end{equation}
For $\Delta >0$, $\Delta=0,$ and  $\Delta < 0$, the equation is 
classified as  hyperbolic, parabolic, and elliptic, respectively. 

It turns out that this classification is related with the \emph{weak energy condition}, 
which ensures that no observer can measure a negative energy density. (It is interesting to
note that the weak- and strong-energy conditions are equivalent in the present case.) 
For the stress-energy tensor given in Eq.~(\ref{T0}), this condition translates into 
\begin{equation}
\rho_0+ \tau_0 > 0 \:\: \Leftrightarrow \:\:  \frac{\lambda}{kL} + 
\frac{1}{2} \left( 1 - (\gamma s ) ^2 \right) > 0.
\label{WEC}
\end{equation}
As a result, the compliance with the weak energy condition is equivalent to 
the equation being hyperbolic (in the whole domain). It is interesting to note 
that in order to comply with causality, one must not assume $\lambda=0$
to model light springs; otherwise, Eq.~(\ref{WEC})  would be violated whenever 
the spring is stretched, i.e. , $\gamma s >1$.

Although the weak energy condition ensures that the differential equation 
describing the spring dynamics has a hyperbolic character, which implies that 
information travels at finite speeds, it does not require such speeds to be 
smaller than the speed of light. For this to be the case, we need to
further constrain the properties of the elastic material composing the spring. 
This is done by analyzing the characteristic curves of Eq.~(\ref{UE}), as 
they are responsible for propagating the information about the initial data. 
By using $u = \partial x/\partial t$ and $s = \partial x/\partial l$, 
Eq.~(\ref{UE}) can be rewritten as a first-order quasilinear system when 
supplemented with the condition $\partial u/\partial l=\partial s/\partial t$. 
Then, by using standard methods, we find  that the slopes (velocities) 
of the characteristic curves $\left(t, l(t)\right)$ in $(t, l)$ coordinates are 
given by  
\begin{equation}
\nu \equiv \frac{dl}{dt}= \pm \frac{1}{\gamma} \frac{1}{\sqrt{{\lambda}/{(kL)}
+\left( 1 - \gamma^2 s^2 \right) /2} \pm \gamma u s}.
\label{CC1}
\end{equation}
 By transforming to the inertial frame $(x^0,x^1)$  using Eq.~(\ref{CT}),  we 
 find that the speed of the characteristics transforms to 
\begin{equation}
\tilde \nu \equiv dx(t, l(t))/dt = \nu s+ u, 
\end{equation}
which, by using Eq.~(\ref{CC1}), can be written as 
\begin{equation}
\tilde{\nu}= \pm \frac{1}{\gamma} \frac{s}{\sqrt{{\lambda}/{(kL)}
+\left( 1 - \gamma^2 s^2 \right)/2} \pm \gamma u s} + u.
\end{equation}
We note that there is an asymmetry in the speed of the characteristic 
curves due to the motion of the points of the spring. To impose that the
 information travels at speeds smaller than the speed of light, it suffices 
 to ensure this in some arbitrary inertial frame due to the relativistic 
 invariance of Eq.~(\ref{UE}). Hence, for each point of the spring, let 
 us take its instantaneous rest frame and put $u=0$ to obtain
\begin{equation}
\tilde \nu = \pm \frac{s}{\sqrt{{\lambda}/{(kL)}+\left( 1 - s^2 \right)/2 }}.
\end{equation}
By demanding  $ |\tilde{\nu} |\leq1$, we find 
\begin{equation}
\frac{\lambda}{kL} -\frac{3s^2-1}{2} \geq 0 
\label{Inequality}
\end{equation}
which constrains the spring (or elastic material) parameters not 
allowing $\lambda/kL$ to be smaller than some quantity that depends 
on the instantaneous local deformation of the spring, $s=\partial x/\partial l$.
We note that the weak energy condition is automatically satisfied once 
inequality~(\ref{Inequality}) is respected, as can be seen from Eq.~(\ref{WEC})
with $\gamma=1$. It is also interesting to note that the weak energy condition 
supplied by $\lambda/(kL)> (-1/2) (1-s^2) + 2 s (1-s)$ implies the dominant 
energy condition. The dominant energy condition is neither implied nor implies 
causality.

\section{Initial and boundary conditions}
\label{sec:boundary conditions}

The initial conditions are determined by giving the position and velocity of each 
spring point at $t=0$. Hence, given two smooth real functions $q(l)$ and $v(l)$~\cite{noteic},
 we have 
\begin{equation}
x(0,l) = q(l)
\label{IC1}
\end{equation}
and
\begin{equation}
\frac{\partial x}{\partial t} (0, l) = v(l).
\label{IC2}
\end{equation}
 
Let us suppose that at $l=0$ the spring is fixed at a massive wall. Then, $x(t,l)$ satisfies 
the boundary condition 
\begin{equation}
x(t,0) = 0
\label{BC1a}
\end{equation}
for any $t \in \mathbb{R}_+.$ We will restrict attention to initial conditions that respect 
the above boundary condition. As a result, $q(l)$  and $v(l)$ will satisfy 
\begin{equation}
q(0) = 0\label{BC1b}
\end{equation}
and
\begin{equation}
v(0) = 0.\label{BC2b}
\end{equation}
The attached body, which has been overlooked until now, enters as a boundary 
condition at $l=L$. The portion of the spring at $l=L$ applies a 3-force $a \tau_0$ 
to the body and since a boost in the direction of the force does not change it, this 
is precisely the 3-force acting on the mass in the wall's frame. Therefore, we can write 
\begin{equation}\label{MassBCa}
\frac{d}{dt} \left( \Gamma M \frac{dX}{dt} \right) = a \tau_0, 
\end{equation}
where $\Gamma(t) \equiv [1-{\left( {dX}/{dt} \right) }^2]^{-1/2}$ 
and $X(t)= x(t,L)$.  By using Eq.~(\ref{HL}) and that 
$Y=kL/a$, we end up with the equation
\begin{equation}\label{MassBCb}
\frac{d}{dt} \left( \Gamma \frac{dX}{dt} \right) 
+ \frac{kL}{M} \left( \Gamma \left. \frac{\partial x}{\partial l} \right|_{(t,L)} - 1 \right) 
= 0.
\end{equation}
This is a more intricate form of boundary condition as it is also a differential
equation and should be solved simultaneously with the spring equation~(\ref{UE}).

The local energy conservation given by Eq.~(\ref{divT}), together with 
the fact that the spring is in Minkowski  spacetime, ensures us that we 
can define a conserved energy in any global inertial frame. The wall's 
frame is particularly interesting as the wall is static and, hence, it does 
no work on the spring. To find the conserved energy, let us integrate
 Eq.~(\ref{divT1}) with respect to $x$ along the whole spring 
 [i.e., from $0$ to $X(t),$ where we are considering Eq.~(\ref{CT})],
\begin{equation}
\int_0^{X(t)} dx  \left( \frac{\partial \rho}{\partial t}+ \frac{\partial g}{\partial x} \right) = 0,
\end{equation}
to obtain
\begin{equation}\label{Epartial}
\frac{d}{d t} \int_0^{X(t)} dx\rho(t, x)- \frac{dX}{dt} \rho(t, X(t)) + g(t, X(t)) -g(t, 0)
= 0.
\end{equation}
By using Eqs.~(\ref{rho}) and~(\ref{g}) together with boundary 
condition~(\ref{BC1a}), we find that  $g-u\rho = u \tau_0$ and $g(t,0) = 0$.
 Hence, we can cast Eq.~(\ref{Epartial}) as
\begin{equation}
\frac{d}{dt} \int_0^{X(t)}d x a \rho(t, x) + U a \tau_0(t,  X(t)) = 0,
\end{equation}
where $U \equiv dX/dt$. Using the boundary condition~(\ref{MassBCa}) 
at $l=L$, the above equation can be rewritten as 
\begin{equation}
\frac{d}{d t} \int_0^{X(t)} dx a \rho(t, x) + U \frac{d}{dt} \left( M \Gamma U \right) = 0,
\end{equation}
which can be easily manipulated to yield
\begin{equation}
\frac{d}{d t} \left[ \int_0^{X(t)} dx a \rho(t, x) +  \Gamma (t) M  \right] = 0.
\end{equation}
As a result, we obtain the conserved energy 
\begin{equation}
{\cal E} \equiv \int_0^{X(t)}  d x \, a \rho(t, x) +  \Gamma (t) M,
\label{energy conservation}
\end{equation}
which is exactly the sum of the energies of the spring and of the body in 
the wall's frame.  

\section{The nonrelativistic regime}
\label{sec:nonrelativistic}

In order to disentangle the relativistic effects from the nonrelativistic ones, 
in this section we analyze the nonrelativistic spring-mass system when 
the spring is massive. Then, in the next section, we will study the relativistic
effects by adding the relativistic corrections to the nonrelativistic solution.

In the nonrelativistic regime, where $u=\partial x/\partial t \ll 1$,  Eq.~(\ref{UE}) reduces to 
\begin{equation}\label{wave}
\frac{1}{\vartheta^2} \frac{\partial^2 x}{\partial t^2} - \frac{\partial^2 x}{\partial l^2} = 0,
\end{equation}
where  
\begin{equation} \label{vartheta}
\vartheta \equiv \sqrt{kL/\lambda}
\end{equation} gives the speed of the perturbations of the above wave equation. In the 
nonrelativistic limit, the boundary conditions~(\ref{BC1a}) and~(\ref{MassBCb}) reduce to 
\begin{equation}\label{nrBC0}
x(t,0)=0
\end{equation} 
and 
\begin{equation}\label{nrBC}
\frac{d^2 X}{dt^2} + \frac{kL}{M} \left(  \left. \frac{\partial x}{\partial l} \right|_{(t, L)} - 1 \right) = 0,
\end{equation}
respectively. 

In the remaining of this section, we will first present a procedure to find a general solution 
of Eq.~(\ref{wave})--with boundary conditions~(\ref{nrBC0}) and~(\ref{nrBC})--in terms 
of the initial conditions~(\ref{IC1}) and~(\ref{IC2}). As it will be discussed, the general 
solution is not suitable to analyze the light-spring case ($m\equiv \lambda L \ll M$) for 
arbitrarily large times. Therefore, we will proceed to show how we can evade this problem to
 find  the motion of the (light) spring and determine how its mass affects the motion of 
 the attached body. 

\subsection{General solution}
\label{sec:nonrelativistic:general solution}

The general solution Eq.~(\ref{wave}) can written as 
\begin{equation}
x(t,l) = f(\vartheta t+l) + h(\vartheta t - l)
\end{equation}
where $f$ and $h$ are two arbitrary smooth functions. Imposing the boundary 
condition~(\ref{nrBC0}) at $l=0$ we find that $h = - f$ and thus 
\begin{equation}\label{nrGS}
x(t ,l) = f(\vartheta t +l) - f(\vartheta t -l).
\end{equation}
By using the initial conditions~(\ref{IC1}) and~(\ref{IC2}) at $t=0$, the function 
$f(z)$, for $z\in [-L,L],$ is determined to be
\begin{equation}\label{f}
f(z) = \left\{ \begin{array}{ll}
\left(   q(z) + {\vartheta}^{-1} \int^{z}_0 dz' v(z') \right)/2  & \mbox{for $z \in [0,L]$} \\
\left(- q(-z) + {\vartheta}^{-1} \int^{-z}_0 dz' v(z') \right)/2  & \mbox{for $z \in [-L,0]$}
\end{array} \right.,
\end{equation}
with which we find $x(t , l)$ for $t \geq0,$ $l\geq 0$ and $\vartheta t + l \leq L$. 
Note that a point near the wall does not know about the existence of the body until 
$t \approx L/\vartheta$, which is the time needed for the information to travel from 
the body, at $l=L$, to $l \approx 0$. 
  
To determine the motion of the body attached to the spring we need to insert
 Eq.~(\ref{nrGS}) into the boundary condition~(\ref{nrBC}). This gives

\begin{equation}
f''(\vartheta t +L) - f''(\vartheta t -L) 
= \frac{\lambda}{M} \left[1- f'(\vartheta t +L) - f'(\vartheta t -L)  \right],
\label{nrBCf}
\end{equation}
where the primes indicate derivatives with respect to the argument.  
By defining $z \equiv \vartheta t + L$  we can rewrite the  above equation as
\begin{equation}\label{nrBC1}
f''(z)  + \frac{\lambda}{M} f'(z) 
=  f''(z-2L) - \frac{\lambda}{M} f'(z-2L) +  \frac{\lambda}{M},
\end{equation}
with which we are able do extend the function $f$ to the domain $[L,3L]$. 
In order to see this, note that whenever $z\in [L,3L]$ the right-hand side of 
Eq.~(\ref{nrBC1}) depends only on the values of $f$ and its derivatives in 
the domain $[-L,L]$, which have already been determined by the initial conditions. 
Hence, we can take $f(L)$ and $f'(L)$ as the initial conditions for Eq.~(\ref{nrBC1}) 
and integrate it from $z=L$ to some $z\in [L, 3L]$ to obtain
\begin{eqnarray}
&&f(z) =f(L) +\left(f(-L) +\frac{M}{\lambda}[f'(L) - f'(-L)]\right)\nonumber \\ 
&&\times \left(1 - e^{-{\lambda (z-L)}/{M}} \right) +\int_L^z  dz' e^{-{\lambda (z-z')}/{M}} 
\nonumber \\
&&\times \left( f'(z'-2L)+ \frac{\lambda}{M}[z' -L-f(z'-2L)]\right).
\end{eqnarray}
Similarly, we can  extend $f$ to the region $[-3L,-L]$ by rewriting Eq.~(\ref{nrBCf})
in terms of $\tilde{z}\equiv \vartheta t - L$ and using $f(-L)$ and $f'(-L)$ as its 
initial conditions when integrating it backwards from $z=-L$ to some $z\in [-3L, -L]$.
This process can be repeated indefinitely  so that at the $N$-th step $f$ is known in
the domain $[-(2N+1)L, (2N+1)L]$, which exactly determines the motion of the body  
up to  some time $t_N = 2NL/\vartheta$. 
 
However, it is difficult to write down an expression for $f$ in terms of a general $N$,
making this kind of solution only useful if we are interested in determining the motion
of the body for short times. Although such a method will be useful when analyzing
the relativistic corrections to nonrelativistic motion of the spring-mass system, 
it is unsuitable to the analysis of the light spring, $m= \lambda L \ll M,$  even 
nonrelativistically.  This is due to the fact that, in this limit, $\vartheta$ gets
very large and many steps will be needed to determine the motion of the body 
up to a reasonable time. For instance, consider the angular frequency for the
ideal (massless) spring, $\omega_0 \equiv \sqrt{k/M}$, and take it as the 
approximate frequency for the light spring case. After a time $t_1 = 2L/\vartheta$ 
the phase will have changed by  

\begin{equation}
\Delta \phi \approx  \omega_0 t_1 = 2L\sqrt{\frac{k}{M}} \sqrt{\frac{\lambda}{kL}} 
= 2 \sqrt{\frac{m}{M}} \ll 2 \pi,
\end{equation}
meaning that $N$ has to be of the order $\sqrt{M/m}$ for one to determine 
just one cycle of oscillation. Next, we will see how this problem can be 
circumvented and we will find the first correction coming from the spring
 mass to the usual harmonic motion of the mass $M$. 

\subsection{The harmonic oscillator case and the first correction from the spring mass}
\label{sec:nonrelativistic:the harmonic oscillator case}

For a massless spring,  $\lambda = 0$, the only admissible configuration is 
the one in which the spring is homogeneously deformed, i.e., each portion
of the spring is deformed by the same ratio. This can be seen directly from 
Eq.~(\ref{wave}), whose solution in this case will be

\begin{equation}
x^{\rm H}(t ,l) = a^{\rm H}(t)l,
\end{equation}
where we have already imposed the boundary condition~(\ref{nrBC0}) at $l=0$. 
Inserting this solution  into Eq.~(\ref{nrBC}) renders the usual equation for the 
harmonic oscillator (apart from the inhomogeneity that comes from the choice 
of the spatial coordinate origin at the wall):
\begin{equation}
\frac{d^2 a^{\rm H}}{dt^2} + \omega_0^2 (a^{\rm H}  - 1)= 0,
\label{NRHO}
\end{equation}
where we recall that 
\begin{equation}\label{omega0}
\omega_0 = \sqrt{\frac{k}{M}}.
\end{equation}
The solution of Eq.~(\ref{NRHO}) is the usual harmonic oscillating one, which allows
us to write $x^{\rm H}(t, l)$ as
\begin{equation}\label{MasslessSol}
x^{\rm H}(t,l) = l \left( 1+ \frac{A}{L} \cos{(\omega_0 t + \varphi)} \right)
\end{equation}
for arbitrary real constants $A$ and $\varphi$. As a result, the motion of the attached 
body is described by the well known behavior: 
\begin{equation}
\label{NRHOS}
X^{\rm H}(t ) = L +A\cos{(\omega_0 t + \varphi)}.
\end{equation}

As we have already shown, however, a massless  spring is unphysical in the relativistic 
regime since it violates the weak energy condition. Furthermore, we are going to see 
that the first relativistic correction over the Newtonian solution is generally smaller than 
the spring's mass correction over a massless spring. Thus, it is essential to study the 
effects of the spring mass on the motion of the attached body. In what follows we will
make the assumption that the spring mass is much smaller than the mass of the body,
i.e., $m\ll M$. Additionally, to simplify the comparison between the nonrelativistic and 
relativistic corrections, we will analyze the special case where all  the spring points oscillate 
harmonically with the same frequency $\omega$. (In Appendix~\ref{App:AppendixB},  we 
analize the case where the mass is stretched and released from rest and obtain 
similar results.) 

Therefore, we will look 
for solutions of Eq.~(\ref{wave}) of the form
\begin{equation}\label{NRharmonicsol}
x(t, l) = b(l) \cos{ (\omega t)} + c(l),
\end{equation}
with $b$ and $c$ being smooth functions defined on the domain $\left[0,L\right]$  satisfying  
\begin{equation}
b(0)=c(0)=0
\label{BCbc0} 
\end{equation}
and
\begin{equation}
\left(\omega_0^2L\left.\frac{db}{dl}\right|_{l=L}-\omega^2b(L)\right)\cos{ (\omega t)}
=-\omega_0^2L\left(\left.\frac{dc}{dl}\right|_{l=L}-1\right)
\label{BCbcL}
\end{equation}
due to the boundary conditions~(\ref{nrBC0}) and~(\ref{nrBC}), respectively. 
Using Eq.~(\ref{NRharmonicsol}) in Eq.~(\ref{wave}) gives
\begin{equation}
\left( \frac{\omega^2}{\vartheta^2} b + \frac{d^2 b}{dl^2} \right) \cos{ (\omega t)} 
+ \frac{d^2 c}{dl^2} = 0,
\end{equation}
which implies that
\begin{equation}\label{bL}
b(l) = A \frac{\sin{(\chi l)}}{\sin{(\chi L)}}
\end{equation}
and 
\begin{equation}\label{cL}
c(l) = B l,
\end{equation}
with $A, B \in \mathbb{R}$ and $\chi\equiv \omega/\vartheta$. Here, we have already 
imposed the boundary condition~(\ref{BCbc0}) on both $b$ and $c$.  Now, by using 
Eqs.~(\ref{bL}) and~(\ref{cL})  in the boundary condition~(\ref{BCbcL}), we arrive at: 
\begin{equation}
A\left[ \omega_0^2 \chi L \cot{(\chi L)}-\omega^2 \right] \cos{ (\omega t)}
+ \omega_0^2L \left[ B-1 \right] = 0,
\end{equation}
from which we obtain that $B=1$ and 
\begin{equation}\label{chiEq}
\chi L = \frac{m}{M} \cot{\! \left(\chi L \right)}.
\end{equation}
This is a transcendental equation for $\chi$ and its solutions, $\chi_n$, can 
be labeled by nonzero integer index $n$ arranged such that
$$
\ldots < \chi_{-2} < \chi_{-1} < 0 < \chi_1 < \chi_2 < \ldots
$$ 
and $\chi_{-n} = -\chi_n$. Hence, for each $\omega_n=\chi_n \vartheta$ satisfying 
Eq.~(\ref{chiEq}), we have that the corresponding solution $x(t, l)$ is given by 
\begin{equation}\label{HarmonicSol}
x(t, l) = l + A \frac{\sin{( \omega_n l/\vartheta)}}{\sin{( \omega_n L/\vartheta)}} \cos{ (\omega_n t)},
\end{equation}
which implies that the motion of the attached body is simply
\begin{equation} \label{nrmoden}
X(t) = L + A  \cos{ (\omega_n t)}.
\end{equation}
Note that this is still a harmonic motion but now with a frequency $\omega_n$ instead of $\omega_0$.

From Eq.~(\ref{chiEq}) we can see that, when $m/M\ll 1$,  $\chi_1L\ll1$ and 
$\chi_{n+1}L\approx n\pi$. Hence, to lowest order, i.e.,  $\tan{\chi_1L}\approx \chi_1L$,  
we find that $\chi_1 L\approx \sqrt{m/M}$ which gives $\omega_1 \approx \omega_0$.  
Approximating the corresponding solution $x(t, l)$ also to the lowest order we find that
\begin{eqnarray}
x(t, l) &=& l + A \frac{\sin{( \omega_1 l/\vartheta)}}{\sin{( \omega_1 L/\vartheta)}} 
\cos{( \omega_1 t)} \nonumber \\
& \approx & l \left( 1 + \frac{A}{L} \cos{( \omega_0 t)}\right).
\end{eqnarray}
As a result, the motion of the body is given by
\begin{equation}
X(t) = L + A  \cos{( \omega_0 t)} ,
\end{equation}
which is exactly the solution for the massless spring.

In order to find the spring's mass correction over the massless spring solution, we will
 need to go to the next order in the Taylor expansion of $\tan (\chi_1L)$. Hence, we 
 write $\tan (\chi_1L) \approx (\chi_1L) + (\chi_1L)^3/3$ and use it in Eq.~(\ref{chiEq}) yielding 
\begin{equation}
\frac{1}{3} (\chi_1L)^4 + (\chi_1 L)^2 - \frac{m}{M} \approx 0,
\end{equation}
which gives, up to order $m^2/M^2$, 
 \begin{eqnarray}
(\chi_1L)^2 \approx \frac{m}{M} - \frac{1}{3} \frac{m^2}{M^2}. 
 \end{eqnarray}
As a result, we find that $\omega_1=\chi_1/\vartheta$ is
\begin{equation}
\omega_1  \approx \omega_0 \left( 1 -  \frac{1}{6}  \frac{m}{M} \right).
\end{equation}
The solution $x(t, l)$, approximated to first order in $m/M$, becomes:
\begin{eqnarray}\label{nr1x}
x(t, l) &=& l + A \frac{\sin{( \omega_1 l/\vartheta)}}{\sin{( \omega_1 L/\vartheta)}} 
\cos{( \omega_1 t)} 
\nonumber \\
& \approx& \!\!\! l + A \frac{l}{L} \left[ 1 + \frac{m}{6M} \left( 1 - \frac{l^2}{L^2} \right) \right]
\cos{\left[ \omega_0 \left(1 - \frac{m}{6M} \right) t \right]},\nonumber\\
\end{eqnarray}
from which we find that the motion of the attached body is given by
\begin{eqnarray}
X(t) &=& L + A \cos{\left[ \omega_0 \left(1 - \frac{m}{6M} \right) t \right]}.
\end{eqnarray}
Hence, up to order $m/M$, the motion of the body is still a simple harmonic oscillation 
with frequency $\omega_1 =\omega_0[1-m/(6M)]$.  If we define $\omega_1\equiv \sqrt{k/M^*}$, 
where $M^*$ is interpreted as the effective mass of the body when the spring is massive, it is easily 
seen that  $M^* = M+m/3$. 

We note that, to first order in $m/M$,  the initial values for Eq.~(\ref{wave}) which give rise to 
solution~(\ref{nr1x}) are
\begin{eqnarray}
x(0,l) &=&  l + A \frac{l}{L} \left[ 1 + \frac{m}{6M} \left( 1 - \frac{l^2}{L^2} \right) \right],\\
\frac{\partial x}{\partial t} (0,l) &=& 0. 
\end{eqnarray}
In contrast, the initial values for the massless spring solution (\ref{MasslessSol}), with $\varphi = 0$, 
are given by 
\begin{eqnarray}\label{inls2a}
x(0,l) &=&  l + A \frac{l}{L}, \\ \label{inls2b}
\frac{\partial x}{\partial t} (0,l)& =& 0.
\end{eqnarray}
Hence, in order to obtain a solution for a light spring perturbatively from the massless spring solution, 
it is mandatory to perturb (and fine tune) the initial conditions of Eq.~(\ref{wave}) accordingly. 
(As it is shown in Appendix~\ref{App:AppendixB}, if one solves  Eq.~(\ref{wave}) for a light spring 
with initial conditions~(\ref{inls2a}) and~(\ref{inls2b}) one will need all modes of oscillation with 
angular frequencies $\omega_n=\chi_n \vartheta$ solving Eq.~(\ref{chiEq}). However, only the 
mode with frequency $\omega_1$ can be perturbatively obtained from the massless 
solution~(\ref{MasslessSol})~\cite{note_pertsol}.)

\section{The relativistic regime}
\label{sec:relativistic}

If the speed of the waves on the spring, $\vartheta$, is comparable with the 
speed of light or the initial conditions are such that the body or the spring 
points eventually move  very fast, the use of the relativistic
equations~(\ref{UE}),~(\ref{BC1a}), and~(\ref{MassBCb}) becomes mandatory.  
Due to their complexity, we will first do a perturbative calculation to determine
the first relativistic correction over the nonrelativistic solution.  
Then, we proceed to solve numerically the full relativistic equations and compare the
results with the nonrelativistic massive case and with the relativistic particle
in the harmonic potential, given by Eq.~(\ref{wronghooke1}).

\subsection{Perturbative analysis}
\label{sec:perturbative}

A rough estimate using the nonrelativistic case suggests that the highest 
speeds achieved by the spring points  are limited by $\vartheta$, as otherwise
Hooke's law would likely be violated. As a simple illustration, consider a very
light spring with the attached body pulled  by a distance $A$ and released
from rest. The body would achieve its maximum speed $V = A\omega_0$
when the spring is instantaneously relaxed. Using Eqs.~(\ref{vartheta})
and~(\ref{omega0}) this can be rewritten as
\begin{equation}
V = \frac{A}{L} \sqrt{\frac{m}{M}} \, \vartheta,
\end{equation}
which is much smaller than $\vartheta$ for $m \ll M$ and $A<L$. One way
to increase this velocity without changing the local properties of the spring
is by taking a lighter body or a longer spring (i.e., with a larger $m$). Therefore,
we would expect the highest maximum speed to be reached in the limit
$M \ll m$, for which, using Eqs.~(\ref{chi}) and~(\ref{NRexpan}) in
Appendix~\ref{App:AppendixB}, we find
\begin{equation}
V =  \frac{A}{L} \frac{\pi}{2}  \vartheta
\end{equation}
in the fundamental mode~\cite{note2}. Thus, the maximum speed 
at which our formulas will be still reliable is roughly limited  by $\mu \vartheta$, 
with $\mu$ being some real constant representing how much the spring can 
be deformed relatively to its relaxed length and still be described by Hooke's law.
A similar analysis can be done in the case where a initial velocity is impressed
on a relaxed spring, instead of stretching and releasing it from rest. Again
we would obtain maximum speeds limited by $\mu \vartheta$.  
Therefore, in the regime where Hooke's law is valid, the relativistic character 
of the motion is determined by the value of  $\mu \vartheta$. For an usual 
coiled spring, $\mu$ can be relatively large (but 
not larger than $1$). However, $\vartheta$ is so small that no visible relativistic 
correction should be expected. On the other hand, for a typical solid elastic material 
replacing the spring, the propagation speed $\vartheta$ can be quite large, but $\mu$
tends to be small (of order $1\%$ or less). Hence, for such cases, it is justified
to treat the relativistic solutions perturbatively with respect to the nonrelativistic
ones. 

Restoring the speed of light, $c$, in the formulas and using $u=\partial x/\partial t $, 
we cast Eq.~(\ref{UE}) as 
\begin{eqnarray}
&&\frac{ \gamma^{2}}{c^2} \left[ \frac{c^2}{\vartheta^2} + \frac{1}{2} 
- \left( \frac{1}{2} + \frac{u^2}{c^2} \right)  
{\left( \gamma\frac{\partial x}{\partial l} \right) \!}^2 \right] \frac{\partial^2 x}{\partial t^2} 
\nonumber \\ 
&&-\frac{\partial^2 x}{\partial l^2} - 2\gamma^{2} \frac{u}{c^2}  
\frac{\partial x}{\partial l} \frac{\partial u}{\partial l } = 0.
\end{eqnarray}
In the case where $u \ll c$, we can retain only the smallest order in $u/c$,
 obtaining the following equation for the spring points:
\begin{eqnarray}
\label{Wavefrc}
\frac{1}{\vartheta^2} \frac{\partial^2 x}{\partial t^2}-\frac{\partial^2 x}{\partial l^2} 
&=& - \frac{1}{c^2} \left[ \frac{u^2}{\vartheta^2} 
+ \frac{1}{2 }-\frac{1}{2 }{\left( \frac{\partial x}{\partial l} \right)}^2 
-2\left(\frac{u}{c}\right)^2{\left( \frac{\partial x}{\partial l} \right)}^2\right] 
\nonumber \\ 
&\times& \frac{\partial^2 x}{\partial t^2} + 2\frac{u}{c^2} \frac{\partial x}{\partial l}  
\frac{\partial u}{\partial l }.
\end{eqnarray}
Using the same approximation in  Eq.~(\ref{MassBCb}), the boundary condition
 at $l=L$ can be written as
\begin{equation}\label{BCfrc}
\frac{dU}{dt}  + \omega_0^2 L  
\left( \left. \frac{\partial x}{\partial l}  \right|_{(t,L)} - 1\right) 
= -\frac{3U^2}{2c^2}\left( \frac{dU}{dt}
+\frac{\omega_0^2 L }{3} \left. \frac{\partial x}{\partial l}\right|_{(t ,L)}\right),
\end{equation}
where we recall that $U(t)=u(t,L)$. We note that the boundary 
condition~(\ref{BC1a}) at $l=0$ remains unchanged. Having obtained the first 
relativistic correction to the nonrelativistic spring equation and boundary condition, 
we again make $c=1$  in what follows. 

Now, we solve Eq.~(\ref{Wavefrc}) perturbatively. For this purpose, 
we assume $\vartheta \ll 1$. Hence, let us write 
\begin{equation}\label{rel_splitting}
x(t,l) = x_m(t,l) + \delta (t,l),
\end{equation}
where $x_m(t, l)$ is a solution of the (massive) nonrelativistic equation 
and  $\delta(t, l)$ is a perturbation such that 
$\delta,$ $\partial\delta/\partial t,$ and $\partial \delta/\partial l$ 
are much smaller than $x_m,$ $\partial x_m/\partial t$, and 
$\partial x_m/\partial l$, respectively.  By using Eq.~(\ref{rel_splitting}) 
in Eq.~(\ref{Wavefrc}), we obtain the following nonhomogeneous wave 
equation for $\delta$:
\begin{eqnarray}\label{PUE}
\frac{1}{\vartheta^2} \frac{\partial^2 \delta}{\partial t^2} 
-  \frac{\partial^2 \delta}{\partial l^2} =  \phi(t, l),
\end{eqnarray}
where 
\begin{eqnarray}
&&\phi(t,l)\equiv 2 \frac{\partial x_m}{\partial l} 
\frac{\partial x_m}{\partial t} \frac{\partial^2 x_m}{\partial l \partial t}
\nonumber \\
&&
- \frac{\partial^2 x_m}{\partial t^2}  \left[ \frac{1}{2} 
\left(1- {\left(  \frac{\partial x_m}{\partial l} \right)}^2\right)
+ \left(\frac{1}{\vartheta^2} \right){\left( \frac{\partial x_m}{\partial t} \right)}^2  \right]. 
\nonumber \\
\end{eqnarray}
We note that the function $\phi$ is defined in the domain
 $[0,L] \times  {\mathbb{R}}^+$. If we now use Eqs.~(\ref{nrBC}) 
 and~(\ref{rel_splitting}) in  Eq.~(\ref{BCfrc}), the boundary condition 
 for $\delta$ at $l=L$ becomes
\begin{equation}\label{PBC}
\left.\left(\frac{\partial^2 \delta}{\partial t^2}  
+ \omega_0^2 L  \frac{\partial \delta}{\partial l}\right)\right|_{(t,L)}  
= \left. \omega_0^2 L{\left( \frac{\partial x_m}{\partial t} \right) \!}^2 
\left( \frac{\partial x_m}{\partial l} - \frac{3}{2} \right)\right|_{(t,L)}.
\end{equation}
As $x(t,0)=x_m(t,0)=0$, the boundary condition for $\delta$ at $l=0$ is 
\begin{equation}
\delta(t, 0)=0.
\end{equation}

By using the retarded Green function
\begin{equation}
G(t,l; t',l') = \frac{\vartheta}{2} \theta \left(t -t' \right) \theta\left(\vartheta (t- t') - |l-l'| \right)
\end{equation}
of the wave operator  
$$
\frac{1}{\vartheta^2} \frac{\partial^2}{\partial t^2}-\frac{\partial^2}{\partial l^2},
$$ 
where $\theta$ is the Heaviside step function, the general solution of Eq.~(\ref{PUE}) can be written as 
\begin{eqnarray}
&&\delta(t,l) = f(\vartheta t + l) + h(\vartheta t - l) 
\nonumber \\
&&+ \frac{\vartheta}{2} \int_{-\infty}^t dt'  \int_{-\infty}^{\infty} dl' \,  
\theta \! \left( \vartheta (t-t') - |l-l'| \right)  \Phi (t',l'). 
\nonumber \\ 
\end{eqnarray}
Here, $f$ and $h$ are arbitrary smooth (or at least $C^2$) functions and $\Phi$ is any extension 
of $\phi$ to the domain $\mathbb{R} \times \mathbb{R}$. A convenient way 
to extend $\phi$ is by taking $\Phi(t, l)$ as an odd function with respect to 
$l$, which is equal to $\phi(t, l)$ in the domain  ${\mathbb{R}}^+ \times [0,L] $ 
and \emph{zero} for $t<0$ or $|l|>L$; i.e., we take 
\begin{equation}
\Phi(t, l) = \theta (t) \left[ \theta(l+L) - \theta(l-L) \right]  \widetilde{\Phi} (t, l),
\end{equation}
where $\widetilde{\Phi}: \mathbb{R} \times \mathbb{R} \to \mathbb{R}$ is any
$C^2$ odd function with respect to $l$ agreeing with $ \phi(t, l)$ in the region
$\mathbb{R}^+ \times [0,L]$.  By choosing 
$f = -h$~\cite{note3}, we can write 
\begin{eqnarray}
\label{delta}
&&\delta(t,l) =  f(\vartheta t + l) -f(\vartheta t - l) 
\nonumber \\
&&\!\!+\frac{\vartheta}{2} \int_0^t dt'  \int_{-L}^{L} dl'  
\theta \left( \vartheta (t-t') - |l-l'| \right)  \Phi (t', l') , \nonumber \\ 
\end{eqnarray}
which, due to our choice for  $\Phi$, automatically satisfies the boundary condition at $l=0$. 

In order to be consistent with our approximations, the initial conditions for 
$\delta$ need to be perturbatively small.  The simplest way to ensure this is by taking them as
\begin{equation}\label{IC_delta1}
\delta(0,l)=0 
\end{equation}
and
\begin{equation}\label{IC_delta2}
\frac{\partial \delta}{\partial t}(0,l)=0. 
\end{equation}
We can now use Eq.~(\ref{delta}) in Eqs.~(\ref{IC_delta1}) and~(\ref{IC_delta2}), 
together with our choice for $\Phi$,  to obtain the conditions
\begin{equation}
f(l)=f(-l) 
\end{equation}
and 
\begin{equation}
 f'(l) = f'(- l) 
\end{equation}
for any $l \in [0,L].$ As a result, $f(z)={\rm const}$ for $z\in[-L,L]$ and we will
set this constant to zero since $\delta$ depends only on the difference of two $f$'s.   
To determine $f(z)$ for $|z| > L$, we will make use of the boundary 
condition~(\ref{PBC}) at $l=L$. To this end, we first use Eq~(\ref{delta}) to evaluate
\begin{eqnarray}\label{partialdelta1}
\frac{\partial \delta}{\partial l} (t,L) &=& f'(\vartheta t + L) + f'(\vartheta t -L) \nonumber \\
&-& \frac{1}{2}  \int_{-L}^{L} dl' \,  \Phi (t - |L-l'|/\vartheta,l') 
\end{eqnarray}
and
\begin{eqnarray}\label{partialdelta2}
&&\frac{1}{\vartheta^2}\frac{\partial^2 \delta}{\partial t^2} (t, L)
=f''(\vartheta t + L) - f''(\vartheta t -L)
\nonumber \\
&&+\frac{1}{2\vartheta}  \int_{-L}^{L} \! dl' \,  
\theta \! \left( t - |L-l'|/\vartheta \right) 
\frac{\partial  \widetilde \Phi}{\partial t} (t - |L-l'|/\vartheta,l')
\nonumber  \\
&& + \frac{1}{2} \Phi(0, L- \vartheta t).
\end{eqnarray}
Next, we use Eqs.~(\ref{partialdelta1}) and~(\ref{partialdelta2}) 
in Eq.~(\ref{PBC}) to obtain the following differential equation for $f$: 
\begin{eqnarray}
\label{rel_f}
&& f''(z+L) + \frac{\lambda}{M}f'(z+L) 
\nonumber \\
&&= f''(z-L) - \frac{\lambda}{M} f'(z-L) + \Psi(z), 
\end{eqnarray}
where $z\equiv \vartheta t \geq L$ and 
\begin{eqnarray}
\Psi(z)\equiv \left.\left( \frac{\partial x_0}{\partial z} \right)^2 \omega_0^2 
L \left( \frac{\partial x_m}{\partial l} - \frac{3}{2} \right) \right|_{l=L} - \frac{1}{2} \Phi(0,L-z) 
\nonumber \\ 
+ \frac{1}{2}  \int_{-L}^{L} \! dl' \,  \theta \! \left( z- |L-l'| \right) 
\left( \omega_0^2 L -  \frac{\partial}{\partial z} \right) 
\widetilde \Phi \left(\frac{z - |L-l'|}{\vartheta},l'\right).
\nonumber \\
\end{eqnarray}
We note that Eq.~(\ref{rel_f}) is very similar to Eq.~(\ref{nrBC1}), analyzed in 
Sec.~\ref{sec:nonrelativistic:general solution}. Hence, the method used to solve 
the differential equation is exactly the same  and will enable us to extend $f$ to 
the domain $[-(2N+1)L,(2N+1)L]$, where $N\in \mathbb{Z}^+$.  By using this 
extension in Eq.~(\ref{delta}), we find the first relativistic correction to the 
nonrelativistic solution $x_m$. In order to find the motion of the attached mass 
$M$, we just use Eq.~(\ref{delta}), with 
\begin{eqnarray}
&& f(z) = 
\int_L^z dz' \left[ f'(z'-2L) -\frac{\lambda}{M} f(z'-2L) \right.
\nonumber \\
&& 
\left. + \int_0^{z'-L} dz'' \Psi (z'') \right] \exp (-\lambda(z-z')/M), \;\;\;  z\geq L,
\nonumber
\end{eqnarray} 
suitably extended, in Eq.~(\ref{rel_splitting}) 
and evaluate $X(t)\equiv x(t,L)= x_m(t,L)+\delta(t,L).$  

\subsection{Numerical analysis}
\label{sec:numerical}

Now, we return to the full relativistic system, governed by Eqs.~(\ref{UE}), (\ref{BC1a}), 
and~(\ref{MassBCb}), and evolve {\it numerically} some initial conditions in order to compare 
the results with both the nonrelativistic massive evolution --- given by Eqs.~(\ref{wave}), (\ref{nrBC0}), 
and (\ref{nrBC}) --- and  the relativistic particle in the harmonic potential --- given by Eq.~(\ref{wronghooke1}).

In both the relativistic and nonrelativistic regimes, the spring-mass system can be characterized 
(after normalizing lengths and time lapses by $L$) by two dimensionless parameters:
$ \lambda L/M = m/M$ and $\lambda /(k L) = \rho_0/Y$ (see Eqs.~(\ref{T0}) and (\ref{HL}) for the 
meaning of $\rho_0$ and $Y$). Recall that causality demands that $\lambda /(kL)$ satisfy inequality~(\ref{Inequality}). 
On the other hand, the relativistic particle in the harmonic potential can be characterized by a single 
dimensionless parameter: $k L^2/M$, which is the ratio of the two parameters of the spring-mass systems.
In what follows, we present the numerical results for two values of each of the spring-mass parameters: 
$\lambda L/M = 2,\,10$ and $\lambda/(kL) = 0.1,\,10$. We also consider two types of initial conditions for 
each combination of parameters.

\begin{figure*}
 \includegraphics[width=\textwidth,height=13cm]{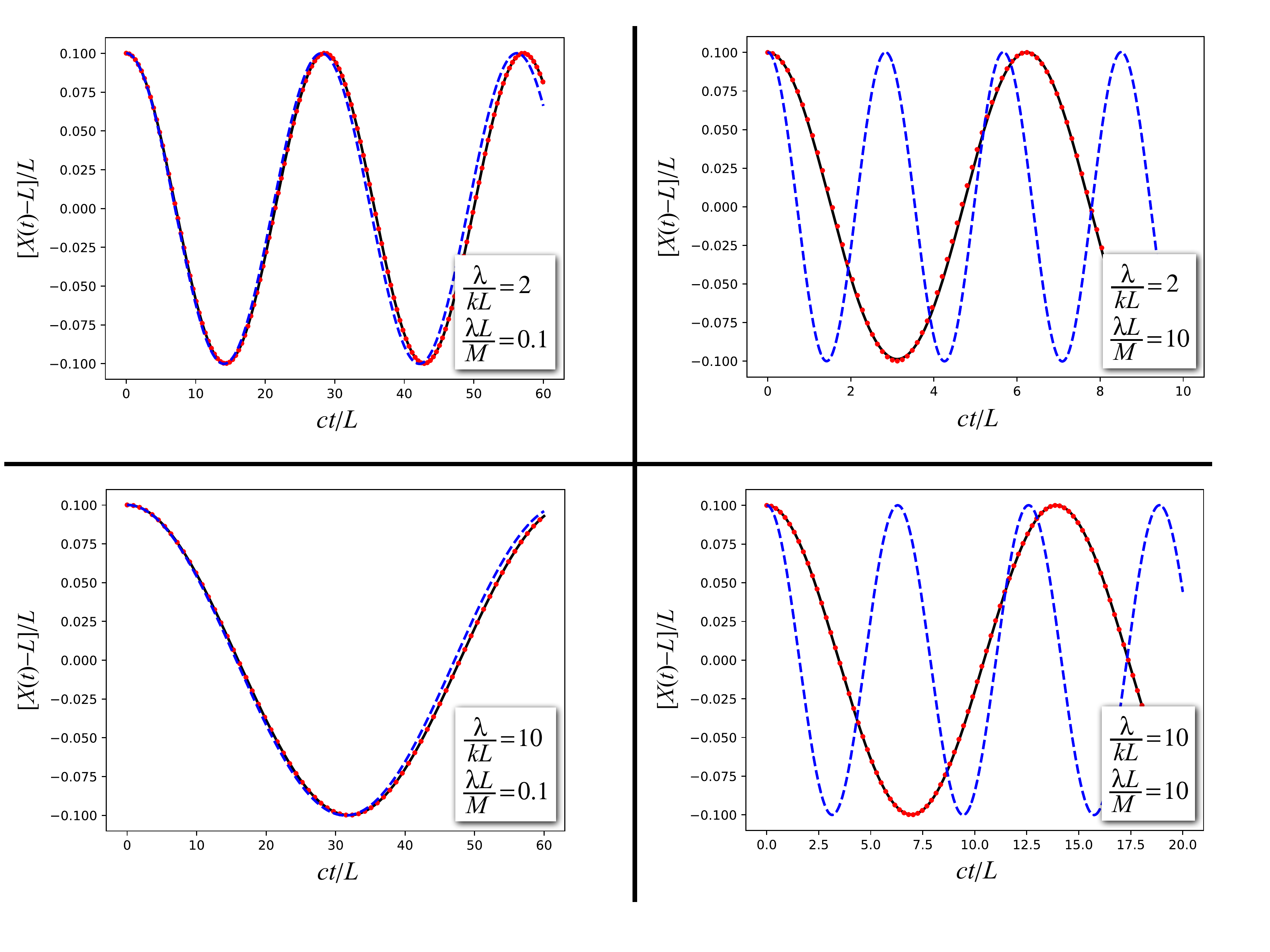}
\caption{Time dependence  of the displacement (with respect to the equilibrium position $X = L$) 
of the mass $M$ attached to the end of the spring, for an initial condition which, in the nonrelativistic 
massive regime, leads to the harmonic oscillation given by Eq.~(\ref{nrmoden}) with $n = 1$ and $A/L = 0.1$. 
The relativistic solution is given by the solid (black) curve, while the (red) dots represent the nonrelativistic 
massive case. For comparison, the (blue) dashed curve gives the position of the relativistic particle in the
 harmonic potential.}
\label{fig:harmIC}
\end{figure*}
In  Fig.~\ref{fig:harmIC}, we plot the time dependence  of the displacement (with respect to the equilibrium 
position $X = L$) of the mass $M$ attached to the end of the spring, for initial conditions which, in the 
nonrelativistic massive regime, lead to the harmonic oscillation given by Eq.~(\ref{nrmoden}) with 
$n = 1$ and $A/L = 0.1$. Both relativistic and nonrelativistic evolutions are plotted, as well as 
 the position of the relativistic particle in the harmonic potential for the corresponding parameter
  $kL^2/M$ (and same initial conditions for its displacement from  the equilibrium position).
We see that, for $m/M \ll 1$, the three systems evolve very similarly. However, for $m/M \gg 1$, 
while both relativistic and nonrelativistic spring-mass evolutions continue to be very similar (for this 
choice of parameters and initial conditions), they differ significantly  from the harmonic-potential one. 
This can be qualitatively understood considering that the  potential energy initially available is converted 
not only to kinectic energy of the mass $M$ but also of the spring elements. Therefore, $M$ moves 
slower when attached to the massive spring than it would in the harmonic potential (with similar initial 
potential energy) --- and this effect is more important the larger the value of $m/M$.
\begin{figure*}
 \includegraphics[width=\textwidth,height=13cm]{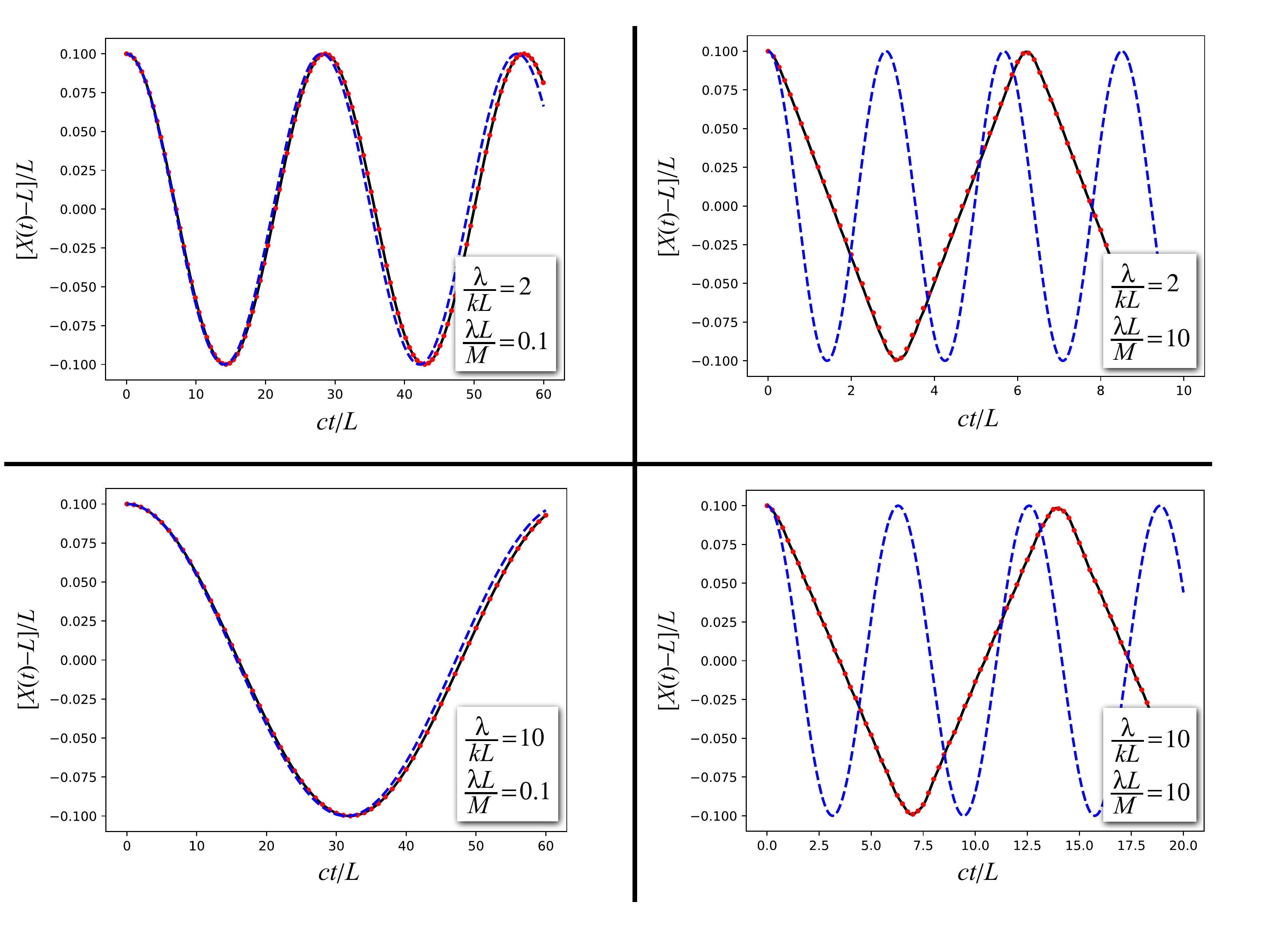}
\caption{Time dependence  of the displacement (with respect to the equilibrium position $X = L$) 
of the mass $M$ attached to the end of the spring, for a spring which is initially  homogeneously 
stretched (by 10\%) and at rest. As in the previous figure, the relativistic solution is given by the 
solid (black) curve, while the (red) dots represent the nonrelativistic massive case. For comparison, 
the (blue) dashed curve gives the position of the relativistic particle in the harmonic potential.}
\label{fig:distIC}
\end{figure*}
In Fig.~\ref{fig:distIC}, we plot the very same quantities as in Fig.~\ref{fig:harmIC}, only changing 
initial conditions to be those of a spring homogeneoulsy stretched (by 10\% of its natural length), at rest.
We see that the previous discussion applies equally well to this case, with the additional point that 
for non-negligible $m/M$, the motion of the mass $M$ attached to the spring is  no longer harmonic. 
This is expected on the grounds that a massive spring has an infinite number of oscillation modes 
and the homogeneously stretched spring excites all of them (see discussion in Appendix~\ref{App:AppendixB}). 
Nonetheless, the nonrelativistic massive-spring evolution continues to approximate very well the relativistic system.
Finally, in Fig.~\ref{fig:harmICRelat} we plot the same quantities as in the previous figures --- with initial conditions
analogous to Fig.~\ref{fig:harmIC} ---, but now
choosing the parameters in such a way that one can perceive a (small) difference between the evolutions of the 
relativistic spring-mass system
and the nonrelativistic one. 
Notwithstanding, even in this ultra-relativistic regime, the relativistic spring-mass
system is farly approximated by the nonrelativistic one. Attempts to
find ultra-relativistic regimes where this
approximation would fail led to numerical solutions which develop 
shock waves (signaled by $s\leq 0$), for which our spring model would not be physically plausible.

\begin{figure}
 \includegraphics[width=8.5cm,height=6.2cm]{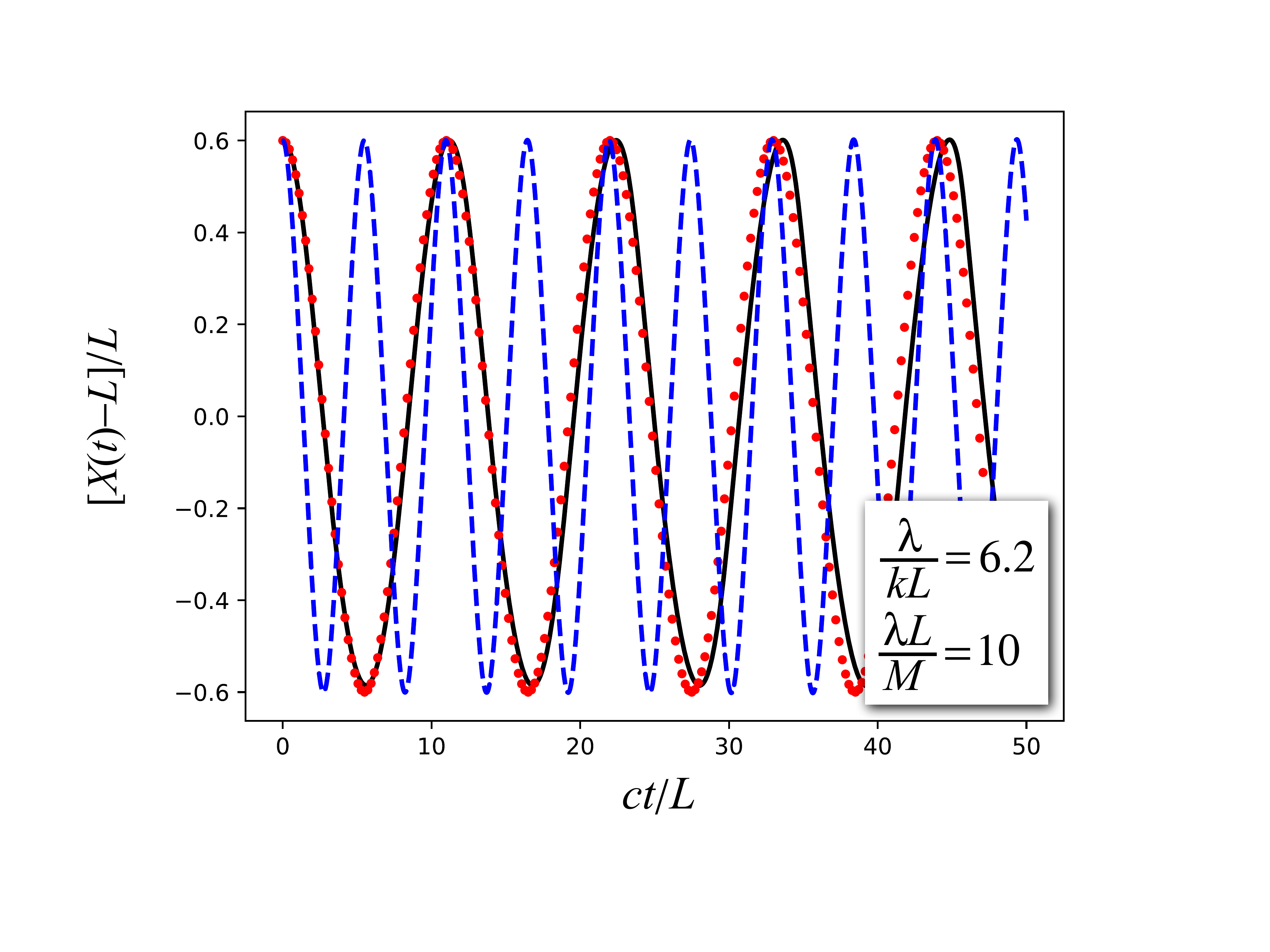}
\caption{Same as Fig.~\ref{fig:harmIC} (with same style code), but now for a particular choice of parameters 
which lead to an ultra-relativistic regime, where a difference between the relativistic
and the nonrelativistic massive spring can be perceived.}
\label{fig:harmICRelat}
\end{figure}

\section{Discussions}
\label{sec:discussions}

We have presented a comprehensive treatment of the relativistic one-dimensional 
Hookean spring-mass  system in Minkowski spacetime. In order to comply with causality, the spring mass 
$m$ must be explicitly taken into account. Interestingly enough, causality is guaranteed if, 
and only if, the weak energy condition (which is equivalent to the strong energy condition in this case)
is verified. In order to see how relativistic contributions impact on the spring-mass system, 
we first analyze the nonrelativistic regime for massive springs and verified that up to order $m/M$, 
the motion of the mass $M$ attached at the spring tip will oscillate harmonically with frequency 
$ \omega_0 [1- m/(6M)]$ up to some time interval of order $(M/m) \omega_0^{-1}$, 
$\omega_0 =\sqrt{k/M}$. Next, we have investigated how the nonrelativistic dynamics is impacted 
by first-order relativistic corrections in $\vartheta/c$. It is interesting to notice that the relativistic dynamics 
was found to be important not only in the regime of high velocities (compared to $c$) 
but also in the regime of high tensions (which may appear in super-hard springs). This can 
be seen from the fact that both $u \ll c$ (small matter velocities) and $\vartheta \ll c$ 
(small sound speed) must be assumed in order to obtain the nonrelativistic equation of 
motion~(\ref{wave}) from the relativistic one, Eq.~(\ref{UE}). Finally, we have performed 
numerical calculations which attest that  the full relativistic spring-mass system is much better 
approximated by the nonrelativistic massive case than by the so-called relativistic harmonic 
oscillator~(\ref{wronghooke1}). Total energy conservation is guaranteed by 
Eq.~(\ref{energy conservation}) and can be used (as it was, in the numerical calculations) to 
monitor the system evolution.

\acknowledgments

R. S., A. L., G. M., and D. V. were fully (R. S.) and 
partially (A. L., G. M., D. V.) supported by S\~ao Paulo Research 
Foundation (FAPESP) under Grants 2015/10373, 2014/26307-8, 2015/22482-2,
and 2013/12165-4, respectively. G. M. was also partially 
supported by Conselho Nacional de Desenvolvimento Cient\'\i fico 
e Tecnol\'ogico (CNPq). G. M would like to acknowledge also 
conversations with Ant\^onio Jos\'e Roque da Silva.
 

\appendix
\section{}
\label{App:Appendix}

Let us assume a charge $q$ with mass $M$ under the influence of 
a fixed external electromagnetic field. The corresponding equation 
of motion will be given by
 \begin{equation}
F^b = M a^b
\label{wronghooketensorial}
\end{equation} 
 with the Lorentz force
(in CGS units):
\begin{equation}
F^b = (q/c) {\cal F}^{b c} u_c, 
\label{Lorentz Eq}
\end{equation}
where $u^b$ and $a^b$ are the charge 4-velocity and 4-acceleration, respectively.
Now, let us assume that the following Faraday tensor approximates the 
electromagnetic field in some region: 
\begin{equation}
{\cal F}^{a b} = (k/q)(x_c X^c) (X^a V^b - V^a X^b),
\label{Faraday}
\end{equation}
where $k$ is a positive constant,
$x^a$ locates spacetime points with respect to some (arbitrarily chosen) spacetime ``origin,''
$V^a$ is a constant
timelike, unit vector field (i.e., $V_a V^a = 1$,
$\nabla_a V^b=0$) which can be associated with the 4-velocity of 
a congruence of inertial Killing observers, while $X^a$ is a 
constant spacelike, unit vector field  orthogonal to $V^a$
(i.e., $X_a X^a= - 1$, $\nabla_a X^b=0$,  and $ X^a V_a =0$). 
Under this assumption, the 
force~(\ref{Lorentz Eq}) acquires the
form 
\begin{equation}
F^a \equiv (k/c) (x_c X^c)
(X^a \;V^b - V^a \; X^b) u_b,
\label{wronghooketensorial3}
\end{equation} 
 which leads to 
Eqs.~(\ref{wronghooke1}) and~(\ref{wronghooke2}), with 
Hooke's 
potential  $\Phi = k (x^a  X_a )^2/2$,
when substituted into Eq.~(\ref{wronghooketensorial}) and, then, projected along $X^a$ and
$V^a$, respectively. It rests to understand to
which electric and magnetic fields Eq.~(\ref{Faraday}) corresponds. 
According to observers of the inertial congruence, this  corresponds 
to the electric  field 
\begin{equation}
E_X = F^{a b} V_a X_b  =  -(k/q) |x^a X_a|
\label{E_X}
\end{equation}
and to no magnetic field. One can easily see that such an electric 
field is found in the interior of a thick plate having constant charge 
density $\rho= - k/(4\pi q)$ and surface linear dimensions much larger than its width.
In this case,  the electric field~(\ref{E_X}) is perpendicular to the plate's surface.
Thus, a charge $q>0$ with mass $M$ moving inside  
such a plate perpendicularly to its surface will oscillate as described by 
Eqs.~(\ref{wronghooke1}) and~(\ref{wronghooke2}), with $(t,x,y,z)$ 
being the Cartesian  coordinates attached to the (preferred) 
laboratory reference frame where the plate lies at rest. Causality 
is not an issue here because the external potential is fixed a 
priori, in contrast to the spring-mass system where the 
corresponding potential depends on the spring state at each moment. 

\section{}
\label{App:AppendixB}

Here, we will determine the dynamics of a (nonrelativistic) massive spring  with 
one end fixed to a inertial wall and the other one attached to a massive body with 
mass $M\gg m$, where we recall that $m=\lambda L$ is the spring's mass.  To this 
end, it will be convenient to define the new function $\tilde x (t, l) = x(t,l)- l$ 
in order to make the system of equations being solved homogeneous. It is easy to see
from Eqs.~(\ref{wave}),~(\ref{nrBC0}), and~(\ref{nrBC}) that $\tilde x$ satisfy the 
nonrelativistic spring equation
\begin{equation}\label{wave2}
\frac{1}{\vartheta^2} \frac{\partial^2 \tilde x}{\partial t^2} - 
\frac{\partial^2 \tilde x}{\partial l^2} = 0
\end{equation}
as well as the boundary conditions $\tilde x(t, 0) = 0$ and 
\begin{equation}\label{nrBCtilde}
\left. \left( \frac{\partial^2 \tilde x}{\partial t^2} +\omega_0^2 L
\frac{\partial \tilde x}{\partial l} \right) \right|_{(t, L)}  = 0
\end{equation}
at $l=0$ and $l=L$, respectively. We note that the boundary condition at $l=L$ is 
now given by a homogeneous  equation, in contrast with Eq.~(\ref{nrBC}).  Next, 
we write the function $\tilde x$ as the product of two independent functions
\begin{equation}\label{separation}
\tilde x(t,l) = \tau(t) \Lambda(l) 
\end{equation}
and insert this into Eq.~(\ref{wave2}) to get
\begin{equation}
\frac{d^2 \Lambda}{dl^2} + \chi^2  \Lambda = 0
\label{Lambda}
\end{equation}
and
\begin{equation}
\frac{d^2 \tau}{dt^2} + \chi^2 \vartheta^2 \tau = 0,
\label{tau}
\end{equation}
where $\chi$ is some real constant. Due to the boundary conditions, we will need 
to consider only the case $\chi \ne 0$ since when $\chi=0$ the corresponding 
solution for $\tilde x$ vanishes. As a result, the general solutions of 
Eqs.~(\ref{Lambda}) and~(\ref{tau}) are given  by 
\begin{equation}\label{Lambdasol}
\Lambda(l) = \sin{(\chi l)}
\end{equation}
and
\begin{equation}
\tau(t) = a \sin{(\chi \vartheta t)} + b \cos{(\chi \vartheta t)},
\end{equation}
respectively, where we have already imposed the boundary condition 
$\tilde x(t,0)=0$ at $l=0.$ To impose the boundary condition at $l=L,$ 
we first use the spring equation (\ref{wave2}) to rewrite Eq.~(\ref{nrBCtilde}) as 
\begin{equation}
{\left. \left(\frac{\partial^2 \tilde x}{\partial l^2}  + \frac{\omega_0^2 L}{\vartheta^2}  
\frac{\partial \tilde x}{\partial l} \right) \right|}_{(t,L)}  = 0.
\end{equation}
By plugging Eq.~(\ref{separation}) in the above equation we find that 
$\Lambda$ satisfies
\begin{equation}\label{LambdaBC}
\left. \left(\frac{d^2 \Lambda}{dl^2}  + \frac{\omega_0^2 L}{\vartheta^2}  
\frac{d \Lambda}{dl}\right)\right|_{l=L}=0
\end{equation}
which, together with Eq.~(\ref{Lambdasol}), gives the relation 
\begin{equation}\label{chi}
\chi L = \frac{m}{M} \cot{(\chi L)}.
\end{equation}
This is a transcendental equation for $\chi$ and its solutions, $\chi_n$, can 
be labeled by nonzero integer index $n$ arranged such that
$$
\ldots < \chi_{-2} < \chi_{-1} < 0 < \chi_1 < \chi_2 < \ldots
$$ 
and $\chi_{-n} = -\chi_n$. Hence, we can retain only the solutions with 
$n>0$ inasmuch as they are completely equivalent to the ones with
 $n< 0.$ As a result, the general solution for the massive spring equation 
 satisfying our boundary conditions is 
\begin{equation}
\label{NRexpan}
x(t,l) = l + \sum_{n=1}^\infty \left[a_n \sin{\left( \chi_n \vartheta t \right)} 
+ b_n \cos{\left( \chi_n \vartheta t \right)} \right]\sin{\left( \chi_n l\right)}.
\end{equation}

Now, the initial conditions~(\ref{IC1}) and~(\ref{IC2}) may be used to determine
 the coefficients $a_n$ and $b_n$ since
\begin{equation}
q(l) = l + \sum_{n=1}^\infty  b_n \sin{\left( \chi_n l\right)}
\end{equation}
and
\begin{equation}
v(l) =\vartheta \sum_{n=1}^\infty   a_n \chi_n \sin{\left( \chi_n l\right)}.
\end{equation}
The problem is that these are expansions in terms of functions that are not 
orthogonal, making it difficult to explicitly solve for the coefficients. 
Nonetheless, the situation is considerably simplified in the light spring case. 
From Eq.~(\ref{chi}) we can see that, when $m/M\ll 1$,  $\chi_1L\ll1$ and 
$\chi_{n+1}L\approx n\pi$. Hence, to lowest order, i.e.,  $\tan{\chi_1L}\approx \chi_1L$,  
we find that $\chi_1 L\approx \sqrt{m/M}$ and thus
\begin{equation}
q(l) = l +b_1  \sqrt{\frac{m}{M}} \frac{l}{L}
+ \sum_{n=1}^\infty b_{n+1}\sin{\left(n \pi \frac{l}{L} \right)} 
\end{equation}
and 
\begin{equation}
v(l) =\frac{\vartheta}{L}\left[ a_1\frac{m}{M}\frac{l}{L} 
+ \sum_{n=1}^\infty n\pi a_{n+1}\sin{\left(n\pi \frac{l}{L}\right)}\right].
\end{equation}

The coefficient $a_1$ and $b_1$ can be easily determined by evaluating 
the above equations at $l=L$, yielding 
\begin{equation}\label{a1}
a_1 = \frac{ML}{m\vartheta}v(L)
\end{equation}
and
\begin{equation}
b_1 =\sqrt{\frac{M}{m}}( q(L) - L), 
\end{equation}
respectively. To determine the coefficients $a_n$ and $b_n$ for $n>1$, 
we just use the orthogonality of the basis functions, $\sin{\left(n \pi \frac{l}{L} \right)},$
to obtain 

\begin{equation}
a_{n+1} = \frac{2}{n\pi \vartheta} \int_0^L dl \left[ v(l) - v(L) \frac{l}{L} \right] 
\sin{\left(n \pi \frac{l}{L} \right)} 
\end{equation}
and
\begin{equation}\label{bn}
b_{n+1} = \frac{2}{L} \int_0^L dl \left[ q(l) - q(L) \frac{l}{L} \right] 
\sin{\left(n \pi \frac{l}{L} \right)}. 
\end{equation}

As a result, the solution describing a very light spring is 
\begin{eqnarray}
\label{GoodMasslessSol}
&&
x(t, l) = l +\sqrt{\frac{m}{M}} \left[ a_1 \sin{(\omega_0 t)} 
+ b_1 \cos{(\omega_0 t)} \right] \frac{l}{L}
\nonumber \\
&&
+ \sum_{n=1}^\infty \sin{\left( n \pi \frac{l}{L}\right)} 
\left[ a_{n+1} \sin{\left( \omega_n t \right)} \
+ b_{n+1} \cos{\left( \omega_n t \right)} \right],
\nonumber \\
\end{eqnarray}
where $\omega_n \equiv n \pi \sqrt{M/m}\, \omega_0$.
and we can see the terms coming from the massless spring's 
equation in the first line, which we call slow waves, 
and in the second line the series representing what we call 
fast waves. These fast waves are responsible to enable arbitrary 
freedom in the choice of initial conditions. We note, however, 
that in this lowest order approximation the motion of the body 
is not affected by these fast waves since $X(t)= x(t,L)$ 
is given by
\begin{equation}
X(t) = L + \left[ a'_1 \sin{(\omega_0 t)} + b'_1 \cos{(\omega_0 t)}  \right],
\end{equation}
where $a'_1=\sqrt{m/M}a_1$ and $b'_1=\sqrt{m/M} b_1,$ which 
we recognize to be the massless spring solution~(\ref{NRHOS}).  
Notwithstanding this, the above procedure will be useful when 
calculating the first  correction to the motion of the body coming 
from the spring mass. 

To determine such correction, let us write 
\begin{equation}\label{NRperturbation}
x(t,l) = x_0(t,l) + \epsilon(t, l)
\end{equation}
where $\epsilon(t, l)$ is a perturbation of order $m/M$ over the 
spring solution  $x_0(t,l)$ given by the right-hand side of 
Eq.~(\ref{GoodMasslessSol}). It should be stressed that even 
though $\epsilon$ is assumed to be a perturbation of small amplitude, 
it will enter the equations with the same importance as the unperturbed 
solution due to the fast waves, which have a substantial second derivative 
in time.

For the sake of simplicity, we will consider the case where the spring is 
held stretched for a while and then released from rest, in which case
\begin{equation}\label{q(l)}
q(l) = l \left( 1 + \frac{A}{L} \right)
\end{equation}
and
\begin{equation}\label{v(l)}
v(l) =0.
\end{equation}
By using the above initial conditions in Eqs.~(\ref{a1})-(\ref{GoodMasslessSol}), 
the unperturbed solution can be written as 
\begin{equation}\label{GoodMasslessSol2}
x_0(t,l) = l \left( 1 + \frac{A}{L} \cos{(\omega_0 t)} \right).
\end{equation}
If we now use Eqs.~(\ref{NRperturbation}) and~(\ref{GoodMasslessSol2}) 
in Eq.~(\ref{wave}), we find that $\epsilon$ satisfies the inhomogeneous equation
\begin{equation}\label{NRIQ}
\frac{1}{\vartheta^2} \frac{\partial^2 \epsilon}{\partial t^2}
 - \frac{\partial^2 \epsilon}{\partial l^2} = \frac{m}{M}\frac{A l}{L^3} \cos{(\omega_0 t)}.
\end{equation}
The boundary conditions and initial conditions for $\epsilon$ are 
straightforward obtained from Eqs.~(\ref{nrBC0}),~(\ref{nrBC}), 
and~(\ref{NRperturbation}) yielding 
\begin{eqnarray}
\epsilon (t, 0) = 0, \;\;\;\; \left.  \left(\frac{\partial^2 \epsilon}{\partial t^2} 
+\omega_0^2 L\frac{\partial \epsilon}{\partial l} \right) \right|_{(t,L)}  = 0&&
\end{eqnarray} 
for the boundary conditions and 
\begin{eqnarray}
\epsilon (0,l) = 0, \;\;\; \left. \frac{\partial \epsilon}{\partial t} \right|_{(0,l)} = 0  &&
\end{eqnarray}
for the initial conditions. 

To solve the above equations, it is convenient to separate $\epsilon$ in two parts:
\begin{equation}\label{deltasplit}
\epsilon(t,l) = \alpha(t,l)+\beta(t,l),
\end{equation}
such that $\alpha$ is a slow-varying solution solving the inhomogeneous 
equation~(\ref{NRIQ}), while $\beta$ will include slow- and fast-varying solutions
solving the homogeneous wave equation
\begin{equation}\label{beta1}
\frac{1}{\vartheta^2} \frac{\partial^2 \beta}{\partial t^2} 
- \frac{\partial^2 \beta}{\partial l^2} = 0.
\end{equation} 
As we have some freedom in choosing the initial and
boundary conditions for $\alpha$ and $\beta$, let us impose $\alpha(t,0)=0$ and 
\begin{equation}
\left. \left( \frac{\partial^2 \alpha}{\partial t^2} 
+\omega_0^2 L\frac{\partial \alpha}{\partial l} \right) \right|_{(t,L)}  = 0.
\end{equation}
(Eventually, we confirm that the solution $\alpha(t, l)$  obtained in this way 
is in agreement with our previous assumption that $\alpha (t, l)$ is slow 
varying.)
This implies that $\beta$ satisfies
\begin{eqnarray}
\beta(t,0) = \epsilon(t,0) - \alpha(t, 0) = 0 \label{beta2}\\
\left.  \left( \frac{\partial^2 \beta}{\partial t^2} 
+\omega_0^2 L \frac{\partial \beta}{\partial l} \right) \right|_{(t, L)}  = 0. 
\label{beta3}
\end{eqnarray}

Since $\alpha$ is a slow-varying function, let us try the following solution 
\begin{equation}\label{alphaSol}
\alpha(t,l) = 
- \frac{A}{6} \frac{m}{M} \frac{l^3}{L^3} \cos{(\omega_0 t)} 
+ a(t) l + b(t)
\end{equation}
for the equation
\begin{equation}
\frac{1}{\vartheta^2} \frac{\partial^2 \alpha}{\partial t^2} - 
\frac{\partial^2 \alpha}{\partial l^2}
=  \frac{m}{M}\frac{A l}{L^3} \cos{(\omega_0 t)},
\label{equation}
\end{equation} 
where $a(t)$ and $b(t)$ are arbitrary smooth functions. 
By imposing the boundary conditions at $l=0$ and $l=L$ we find 
that $b(t)=0$ and $a(t)$ satisfies
\begin{equation}
\frac{d^2 a}{dt^2} + \omega_0^2 a 
= \frac{A}{3L} \frac{m}{M} \omega_0^2 \cos{(\omega_0 t)}.
\end{equation}
We will take $a(t)$ to be the following particular solution of the above equation:
\begin{equation}
a(t)  = \frac{A}{6L} \frac{m}{M} \omega_0 t \sin{(\omega_0 t)},
\end{equation}
which enable us to cast a particular solution of $\alpha$ as
\begin{equation}
\label{alpha}
\alpha(t,l) = - \frac{Al^3}{6L^3}\frac{m}{M}  \cos{(\omega_0 t)} 
+  \frac{Al}{6L} \frac{m}{M} \omega_0 t \sin{(\omega_0 t)}.
\end{equation}

Now, we demand
\begin{equation}
\omega_0 t \ll \frac{M}{m}
\end{equation}
to guarantee that Eq.~(\ref{alpha}) approximates a solution of Eq.~(\ref{equation}).
As a result, our perturbed solution is expected to be accurate only up to a 
certain time interval. However, since $M/m \gg 1$, many cycles of oscillation 
will take place during this time interval.

Now, we note from Eqs.~(\ref{beta1})-(\ref{beta3}) that the problem of finding 
the solution for $\beta$ is exactly the same problem we have already solved for the 
light spring case  [see Eqs.~(\ref{wave2})-(\ref{nrBCtilde})] with initial conditions
\begin{equation}
\beta(0,l) =  \epsilon(0,l) - \alpha(0,l) =\frac{A}{6} \frac{m}{M} \frac{l^3}{L^3} \omega_0^2 
\end{equation}
and
\begin{equation}
\frac{\partial \beta}{\partial t}(0,l) =  
\frac{\partial \epsilon}{\partial t}(0,l) - \frac{\partial \alpha}{\partial t}(0,l) = 0.
\end{equation}
Hence, we can use Eq.~(\ref{GoodMasslessSol}) together with 
Eqs.~(\ref{a1})-(\ref{bn}) to find that $\beta$  is given by
\begin{eqnarray}\label{betasol}
&&\beta(t,l)
=A\frac{m}{M}\left[ \frac{l}{6L} \cos{(\omega_0 t)}\right. 
\nonumber \\
&&+\left.\frac{2}{\pi^3} \sum_{n=1}^\infty \frac{(-1)^n}{n^3} 
\sin{\left( n \pi \frac{l}{L}\right)} \cos{\left( \omega_n t \right)}\right].
\end{eqnarray}
As a result, using Eqs.~(\ref{GoodMasslessSol2}),~(\ref{deltasplit}),~(\ref{alpha}), 
and~(\ref{betasol}) in Eq.~(\ref{NRperturbation}), we find that the solution for the 
spring equation corrected to first order in $m/M$ and satisfying the 
boundary conditions~(\ref{nrBC0})-(\ref{nrBC}) as well as initial 
conditions~(\ref{q(l)})-(\ref{v(l)}) is 
\begin{eqnarray}
&&x(t, l)= l\left(1 + \frac{A}{L} \cos{(\omega_0 t)} \right) 
\nonumber \\
&+&\frac{A}{6}\frac{m}{M}\frac{l}{L}\left[\left(1-\frac{l^2}{L^2}\right)
\cos{\left(\omega_0 t\right)} + \omega_0 t \sin{(\omega_0 t)}\right] 
\nonumber \\
&+& A \frac{m}{M}\frac{2}{\pi^3} \sum_{n=1}^\infty \frac{(-1)^n}{n^3} 
\sin{\left( n \pi \frac{l}{L}\right)} \cos{\left( \omega_n t \right)}.
\nonumber \\
\end{eqnarray}


\begin{thebibliography}{}
\bibitem{note1} We note that nonlongitudinal pressures and shear stresses, if any, 
 should not affect our results because of the one-dimensional feature of our problem.
\bibitem{noteic}
On physical grounds, $q(l)$ must be a monotonically increasing function of $l$, while
$|v(l)|<1$.
\bibitem{note_pertsol}
The missing of all but one solution unveils the non-perturbative character of the light spring case over 
the massless spring one. This is actually reasonable as perturbations travel extremely fast on a very 
light spring. This can be easily seen by noting that the massless spring equation cannot  deal with situations 
where the spring is released from an inhomogeneous configurations.  The massive spring equation, however, 
will be able to handle  this by superposing to the slowly oscillating modes some very fast oscillating ones.
\bibitem{note2} The higher modes exhibit maximum velocities of similar order, 
if one assumes small  deformations.
\bibitem{note3} We note that this boundary condition demands $f(x)= -g(x)$ 
for $x \ge 0$, but it says nothing about the relation between $f$ and $h$ for 
$x<0$. We can, however, extend the validity of the condition $f(x)=-h(x)$  to
the whole domain provided that it is still possible to satisfy the boundary condition 
at $l=L$ and allow arbitrary initial conditions.
\end{thebibliography}
\end{document}